\documentclass[a4paper,11pt]{article}
\pdfoutput=1 

\usepackage{jheppub} 

\usepackage[T1]{fontenc} 

\usepackage{slashed}
\usepackage[utf8]{inputenc}

\def\to{\rightarrow}
\def\ord{\mathcal{O}}
\def\TeV{~{\mbox{TeV}}}

\def\GeV{~{\mbox{GeV}}}

\def\mL{\mathcal{L}}

\def\tree{\text{tree}}
\def\onel{\text{1-loop}}
\def\ts{\text{s}}
\def\tns{\text{ns}}

\def\TH{\text{TH}}
\def\RG{\text{RG}}
\def\TOT{\text{TOT}}
\def\SM{\text{SM}}

\title{\boldmath Simplest Little Higgs Revisited: Hidden Mass Relation, Unitarity and Naturalness}

\author[a,d,e]{Kingman Cheung,}
\author[b]{Shi-Ping He,}
\author[c]{Ying-nan Mao,}
\author[a]{Chen Zhang,}
\author[b]{Yang Zhou}

\affiliation[a]{Physics Division, National Center for Theoretical Sciences, Hsinchu, Taiwan 300}
\affiliation[b]{Institute of Theoretical Physics \& State Key Laboratory of Nuclear Physics and Technology, Peking University, Beijing 100871, China}
\affiliation[c]{ Center for Future High Energy Physics \& Theoretical Physics Division,
Institute of High Energy Physics, Chinese Academy of Sciences, Beijing 100049, China}
\affiliation[d]{Department of Physics, National Tsing Hua University, Hsinchu 300, Taiwan}
\affiliation[e]{Division of Quantum Phases and Devices, School of Physics,
Konkuk University, Seoul 143-701, Republic of Korea}

\emailAdd{cheung@phys.nthu.edu.tw}
\emailAdd{sphe@pku.edu.cn}
\emailAdd{maoyn@ihep.ac.cn}
\emailAdd{czhang@cts.nthu.edu.tw}
\emailAdd{yangzhou9103@gmail.com}

\preprint{NCTS-PH/1803}

\abstract{We analyze the scalar potential of the Simplest Little Higgs (SLH)
model in an approach consistent with the spirit of continuum effective
field theory (CEFT). By requiring correct electroweak symmetry breaking (EWSB)
with the $125\GeV$ Higgs boson, we are able to derive a relation between
the pseudo-axion mass $m_\eta$ and the heavy top mass $m_T$, which serves
as a crucial test of the SLH mechanism. By requiring $m_\eta^2>0$ an upper
bound on $m_T$ can be obtained for any fixed SLH global symmetry breaking
scale $f$. We also point out that an absolute upper bound on $f$ can be obtained
by imposing partial wave unitarity constraint, which in turn leads to
absolute upper bounds of $m_T\lesssim 19\TeV, m_\eta\lesssim 1.5\TeV$ and $m_{Z'}\lesssim 48\TeV$.
We present the allowed region in the three-dimensional parameter space
characterized by $f,t_\beta,m_T$, taking into account the requirement of
valid EWSB and the constraint from perturbative unitarity. We also
propose a strategy of analyzing the fine-tuning problem consistent with the
spirit of CEFT and apply it to the SLH. We suggest that the scalar potential
and fine-tuning analysis strategies adopted here should also be applicable
to a wide class of Little Higgs and Twin Higgs models, which may reveal
interesting relations as crucial tests of the related EWSB mechanism and
provide a new perspective on assessing their degree of fine-tuning.

}

\begin{document}
\maketitle
\flushbottom

\section{Introduction}
\label{sec:intro}

The discovery of the $125\GeV$ Higgs-like particle~\cite{Aad:2012tfa,Chatrchyan:2012xdj} is undoubtedly a great
success of the Standard Model (SM) in which the electroweak symmetry
breaking (EWSB) is achieved via the non-zero vacuum expectation value
associated with a single $SU(2)_L$ doublet Higgs field. Nevertheless,
there is no \textit{a priori} reason to believe that the EWSB must be
realized in the minimal manner dictated by the SM. There are in fact
compelling signs that physics beyond the SM (BSM) should
exist to account for issues like dark matter, neutrino mass and
oscillation, baryon asymmetry of the universe, etc. In a general setting,
the new physics responsible for the explanation of these issues would
interact with the SM Higgs field such that the Higgs mass becomes
radiatively unstable\footnote{This is the Higgs mass naturalness or fine-tuning
problem, which we refer the reader to ref.~\cite{Giudice:2013nak,Giudice:2008bi}
and references therein for representative discussion in the previous
literature. In this work we do not distinguish semantically between ``naturalness problem''
and ``fine-tuning problem'' of the Higgs mass. The detailed meaning of
the Higgs mass naturalness problem is explained in Section~\ref{sec:nislh}.}. It is therefore preferable that some mechanism
should exist to stabilize the Higgs mass. Very often these stabilizing mechanisms
would require modification of the minimal EWSB mechanism realized
by one single $SU(2)_L$ doublet Higgs field. Also, such modification should be related
to a scale not much higher than the electroweak scale so that the
stabilizing mechanism itself does not introduce a severe fine-tuning
problem.

One popular candidate of such stabilizing mechanisms is weak scale
supersymmetry (SUSY), which is representative of weakly-coupled extensions
of the SM. Compared to scenarios which invoke strong dynamics,
SUSY extensions are more predictive in terms of
calculability. On the other hand, SUSY entails the introduction of
superpartners for every SM particle, and hence nearly a doubling
of degrees of freedom in the theory. None of these superpartners
have been observed so far. Also, a large number of new parameters
associated with these new degrees of freedom are introduced,
making the model quite complicated. It is therefore very desirable
if there are other weakly-coupled theories which could stabilize
the Higgs mass and at the same time require less degrees of
freedom with simpler theoretical construction.

One interesting model building option consistent with this line
of thinking is to use the Little Higgs mechanism~\cite{ArkaniHamed:2001nc,
ArkaniHamed:2002pa,ArkaniHamed:2002qx,ArkaniHamed:2002qy}\footnote{For early
reviews, see ref.~\cite{Schmaltz:2005ky,Perelstein:2005ka}.}. The
essential ingredient of the Little Higgs mechanism is collective
symmetry breaking (CSB). In CSB, the Higgs boson is realized as a
Nambu-Goldstone boson (NGB) of some global symmetry breaking. However,
the global symmetry is also explicitly broken in such a manner
that at least two operators are needed at the same time to break enough symmetry
so that the Higgs ceases to be an exact NGB. Because more operators
are needed to renormalize the Higgs mass, the radiative stability
of the theory is improved. Enlargement of gauge group
is generally required for the implementation of the Little Higgs
mechanism. One of the simplest possibilities is the Simplest Little Higgs
(SLH) model~\cite{Kaplan:2003uc,Schmaltz:2004de}, in which the electroweak
gauge group is enlarged to $SU(3)_L\times U(1)_X$. Accordingly,
two scalar triplets are needed, the vacuum expectation values of
which leads to the following spontaneous global symmetry
breaking pattern:
\begin{equation}
[SU(3)_1\times U(1)_1]\times[SU(3)_2\times U(1)_2]
\to[SU(2)_1\times U(1)_1]\times[SU(2)_2\times U(1)_2]
\label{eq:gsb}
\end{equation}
The global symmetry is also explicitly broken by
gauge and Yukawa interactions, but in a collective
manner to improve the radiative stability of the
model. The $125\GeV$ Higgs boson is supposed to be
one of the pseudo-Nambu-Goldstone bosons of the
global symmetry breaking in Eq.~\eqref{eq:gsb}. The
particle content is quite economical. Especially,
for the scalar sector, with the radial modes integrated
out, there is only one physical degree of freedom
left (usually referred to as the ``pseudo-axion''~\cite{Kilian:2004pp,Kilian:2006eh})
besides the $125\GeV$ Higgs boson.

When it comes to the extraction of EWSB predictions
in the SLH, there are two approaches
adopted in the literature\footnote{These two approaches
are not peculiar to the study of the SLH. They have been widely adopted
for many Little Higgs and Twin Higgs models as well.}. The first approach
is to calculate under the assumption of ``No large
Direct Contribution to the scalar potetial from
the physics at the Cutoff''~\cite{Schmaltz:2004de,Reuter:2012sd}, which will
be abbreviated as the ``NDCC assumption'' in the rest of the paper.
In this approach, the tree-level scalar potential
is assumed to vanish (except for a technically natural
$\mu$ term which gives mass to the pseudo-axion).
The one-loop scalar effective potetial is generated by
Yukawa and gauge interactions, triggering EWSB and
making the Higgs boson massive. The divergent loop
momentum integral is assumed to be cut off at the
naive dimensional analysis (NDA)~\cite{Manohar:1983md} cutoff of the associated
nonlinear sigma model. The second approach
is to simply abandon the NDCC assumption
and treat the associated model parameters as effectively
free parameters~\cite{Han:2013ic}.

Both approaches mentioned above have significant
drawbacks if we attempt to derive quantitative predictions
from the SLH. In the first approach, the regularization
cutoff encountered in one-loop effective potential
calculations is invested with
a physical meaning, rather than being treated via the
standard renormalization procedure discussed in quantum
field theory (QFT) textbooks~\cite{Peskin:1995ev} and the original Coleman-Weinberg
paper~\cite{Coleman:1973jx}\footnote{We note that there might be
an interpretational ambiguity about the first approach. One could choose
not to interpret this approach as imparting physical meaning to the
regularization cutoff without renormalization. Instead, one might
interpret the cutoff (usually denoted as $\Lambda$) as the renormalization scale~\cite{Casas:2005ev}.
However, many papers (e.g.~\cite{ArkaniHamed:2002qy,Schmaltz:2004de,Chacko:2005pe}) retain a $\Lambda^2$ term in the
Coleman-Weinberg potential, and demonstrate the cancellation of
quadratic divergence by showing its coefficient vanishes in the
considered model. This is at least formally in conflict with the interpretation of
$\Lambda$ as a renormalization scale in a mass-independent renormalization scheme. We will discuss
this alternative interpretation further in Section~\ref{sec:hmr} and
Section~\ref{sec:nislh}.}. The practice
of imparting a physical meaning to the regularization cutoff
could be somewhat misleading on certain occasions~\cite{Burgess:1992gx}
. Therefore it is always desirable that the relevant
problems be treated in a more rigorous and solid manner
with a clear conceptual foundation. Moreover, if we stick
to the first kind of approach (with its conceptual foundation
put aside for the moment),we would have difficulty in
determining the cutoff value to be used. In the SLH literature
like ref.~\cite{Schmaltz:2004de}, a cutoff value of $4\pi f_1$
is used where $f_1$ denotes the smaller one of the
vacuum expectation values of the two scalar triplets. This
value comes from the NDA which only gives a qualitative rather
than quantitative estimate of the scale up to which the
nonlinear sigma model is expected to be valid. If we consider the requirement of perturbative
unitarity, then usually the cutoff value is much smaller
than the NDA estimate~\cite{Chang:2003vs}. Therefore the results
obtained by plugging in any specific cutoff value cannot be
taken too seriously and we would have no idea about the
associated uncertainties. A further objection is that there
seems to be no \textit{a priori} reason to believe that
there is no large direct contribution from the physics
at the cutoff, and explanations are needed to clarify
what is meant exactly by ``large''.

In the second approach, as adopted in ref.~\cite{Han:2013ic},
although the ad hoc assumption used in the first approach is abandoned,
the parameters related to the EWSB in the SLH are all treated
free parameters which can vary independently. Therefore, the predictivity
of the SLH in terms of its EWSB is lost to a large extent. As we will
show in the following sections, even if we allow direct contribution
to the scalar potential from the physics at the cutoff, there is
still an important mass relation which connects various parameters
of the model dictated by the requirement of correct EWSB with
the $125\GeV$ Higgs boson.

In this paper we would argue that it is both possible and preferable to
adopt an approach consistent with the spirit of continuum effective
field theory (CEFT), which leads to clear and calculable predictions
regarding the EWSB in the SLH. Here, the meaning of CEFT deserves
some remarks. In physics literature, an effective field theory (EFT)
can be discussed in two related but distinct frameworks~\cite{Georgi:1994qn}: one is the
CEFT, and the other is the Wilsonian EFT (WEFT). In the Wilsonian approach,
there is indeed an intrinsic UV cutoff associated with the theory;
this is in contrast to the continuum approach, in which the UV
cutoff only appears in the regularization and should be removed
after renormalization. In this regard, the UV divergences/cutoffs
encountered in the continuum approach can be thought of as \textit{formal}
infinities/cutoffs. The connection between WEFT and CEFT is somewhat
subtle and much confusion could arise from conflating them~\cite{Schwartz:2013pla}.
It should be made quite clear that in the usual phenomenological studies
of both SM and BSM physics based on perturbative quantum field theory,
if without specific declaration, the theoretical framework on which
the calculations are based is CEFT rather than WEFT. This is because
CEFT allows the use of mass-independent renormalization schemes
which are very convenient and facilitate an easy and systematic
construction of the perturbative expansion. On the other hand,
it is awkward to employ WEFT for ordinary phenomenological studies.

With the above in mind, we perform an analysis of the SLH scalar potential
in the CEFT approach. We explicitly write down the scalar quartic
term required by the renormalization procedure without any assumption
on the contribution from the physics at the cutoff. This does not
make the EWSB prediction in the SLH completely arbitrary because
the renormalization is constrained by the symmetry of theory.
Minimization of the scalar effective potential up to one-loop level
is supposed to yield an electroweak vacuum expectation value
and a Higgs mass consistent with experiments. As we will show,
these requirements lead to an interesting mass relation between
the pseudo-axion mass $m_\eta$ and the heavy top mass $m_T$, which serves as
a crucial test of the SLH mechanism. Due to the anti-correlation between
$m_\eta$ and $m_T$, requiring $m^2_\eta>0$ leads
to an upper bound on $m_T$ for any given SLH global symmetry breaking scale
$f\equiv\sqrt{f_1^2+f_2^2}$, where $f_1,f_2$ denote the vacuum expectation
values of the two scalar triplets before EWSB, respectively.
Another prediction of the EWSB in the SLH is that the minimal
value of the ratio between the two scalar triplets $t_\beta\equiv\frac{f_2}{f_1}$
(assuming $f_2\geq f_1$ for the moment) is expected to increase
with the increase of $f$. We note that the heavy gauge boson
masses in the SLH is mainly determined by the overall scale $f$ while
the NDA/unitarity cutoff is supposed to be determined by the smaller
one of the two scalar vacuum expectation values. This implies that
a too large $f$ will push the heavy gauge boson masses into
a region where perturbation theory might not be reliable. In this work
we require all particle masses in the low energy theory do not
exceed the perturbative unitarity bound derived for the SLH nonlinear
sigma fields. Consequently, we are able to obtain absolute upper
bounds on the scale $f$ and all the relevant particle masses in the
theory with which a self-contained EFT for SLH below its unitarity cutoff
can be established.

A further advantage of the CEFT approach is that it automatically
offers a clear and convenient framework for a quantitative investigation
of the naturalness problem. The physical content of a quantum field
theory can be formulated independent of the regularization cutoff
and the naturalness problem should be formulated using only physical,
renormalized quantities. The real issue of the naturalness problem
is radiative stability. We note that in CEFT, the radiative structure of the theory
is embodied in its renormalization group equations (RGE) in a mass-independent
renormalization scheme, keeping in mind that when going below heavy
particle thresholds, we need to put in by hand the Appelquist-Carazzone
decoupling~\cite{Appelquist:1974tg} by integrating out heavy degrees of freedom and shifting
to a low energy EFT suitable for the description of the low energy
phenomena~\cite{Georgi:1994qn}. Within this picture, two sources
of fine-tuning can be easily identified: one is related to
the RGE evolution between thresholds, and the other is related
to the matching onto a low energy EFT when crossing thresholds.
Based on these considerations, we explicitly calculate these
two kinds of fine-tuning for the case of the SLH, using results
obtained from our scalar potential analysis. Furthermore, it is
possible to combine the two fine-tuning measures and obtain a
total fine-tuning for the SLH which is a measure of how sensitive
the electroweak scale parameters are to the variation of the parameters
defined at the unitarity cutoff of the theory. We also clarify
the connection between our fine-tuning definition and
some conventionally adopted fine-tuning definitions used for the SLH in the literature.

The paper is organized as follows. In Section~\ref{sec:slh}, we
introduce the basic setup of the SLH model, including the field content
and the Lagrangian. Section~\ref{sec:hmr} analyzes the SLH
scalar potential in an approach consistent with the spirit of CEFT
with the deriviation of the $m_\eta-m_T$ mass relation and characterization
of the allowed parameter space consistent with valid EWSB and unitarity.
Section~\ref{sec:nislh} clarifies the notion of
naturalness based on the CEFT picture with a quantitative illustration
for the case of the SLH. In Section~\ref{sec:dnc}
we present our discussion and conclusion.


\section{The Simplest Little Higgs}
\label{sec:slh}

In the SLH, the electroweak gauge group is enlarged to $SU(3)_L\times U(1)_X$.
Two scalar triplets $\Phi_1,\Phi_2$ are introduced as nonlinear
sigma fields and parameterized in the following manner
to realize the spontaneous global symmetry breaking
pattern in Eq.~\eqref{eq:gsb}
\begin{align}
\Phi_1=\exp\left(\frac{i\Theta'}{f}\right)
\exp\left(\frac{it_\beta\Theta}{f}\right)
\begin{pmatrix}
0 \\ 0 \\ fc_\beta
\end{pmatrix} \label{eq:phi1} \\
\Phi_2=\exp\left(\frac{i\Theta'}{f}\right)
\exp\left(-\frac{i\Theta}{ft_\beta}\right)
\begin{pmatrix}
0 \\ 0 \\ fs_\beta
\end{pmatrix} \label{eq:phi2}
\end{align}
Here we have introduced the shorthand notation
$s_\beta\equiv\sin\beta,c_\beta\equiv\cos\beta,t_\beta\equiv\tan\beta$.
The Goldstone decay constant $f$ is supposed to be
at least a few$\TeV$. $\Theta$ and $\Theta'$ are $3\times 3$ matrix fields,
parameterized as
\begin{align}
\Theta=\frac{\eta}{\sqrt{2}}+
\begin{pmatrix}
\textbf{0}_{2\times 2} & h \\
h^\dagger & 0
\end{pmatrix},\quad
\Theta'=\frac{\zeta}{\sqrt{2}}+
\begin{pmatrix}
\textbf{0}_{2\times 2} & k \\
k^\dagger & 0
\end{pmatrix}
\label{eq:theta}
\end{align}
where $\eta$ is the physical pseudo-axion discussed in literature~\cite{Kilian:2004pp,Kilian:2006eh}
,and $h$ and $k$ are parameterized as ($v$ denotes
the vacuum expectation value of the Higgs doublet)
\begin{align}
h & =\begin{pmatrix} h^0 \\ h^- \end{pmatrix},\quad
h^0=\frac{1}{\sqrt{2}}(v+H-i\chi) \\
k & =\begin{pmatrix} k^0 \\ k^- \end{pmatrix},\quad
k^0=\frac{1}{\sqrt{2}}(\sigma-i\omega)
\end{align}
For future convenience, we introduce the notation
\begin{equation}
\hat{h}\equiv(h^\dagger h)^{1/2}
\end{equation}
Some remarks about the above parametrization are in order.
Firstly, there is considerable freedom in
parameterizing the scalar triplets $\Phi_1$ and $\Phi_2$.
For instance, in Eq.~\eqref{eq:phi1} and Eq.~\eqref{eq:phi2} we have
adopted a double exponential parametrization. Also,
in Eq.~\eqref{eq:theta} we use the identity matrix
as the generator for the $\eta,\zeta$ fields. It is
certainly legitimate to use instead a single
exponential parametrization, and/or some other appropriate matrix
like $T^8\equiv\frac{\lambda^8}{2}$ ($\lambda^8$
denotes the eighth Gell-Mann matrix) for the $\eta,\zeta$
generator. These different parametrizations are mathematically
related by field redefinition and are thus physically
equivalent. Nevertheless, using the identity matrix
as the generator for $\eta,\zeta$ fields simplifies
the calculation, and as pointed out in ref.~\cite{He:2017jjx},
the double exponential parametrization does not
induce mixing of $\eta$ with unphysical Goldstones
in the term responsible for $\eta$ mass. Secondly,
we have assumed that among various Goldstone components,
only the real part of $h^0$ may acquire a non-zero
vacuum expectation value. Especially, the $\eta$ field
has zero vacuum expectation value and therefore CP
is not spontaneously broken\footnote{We refer the reader
to ref.~\cite{Mao:2017hpp} for a previous
paper on the phenomenology of the spontaneous CP-violating SLH.}. Such a
vacuum configuration can be obtained by minimization
of the scalar effective potential, as will be demonstrated
in Section~\ref{sec:hmr}.

In the SLH, under the full gauge group $SU(3)_C\times SU(3)_L\times U(1)_X$,
$\Phi_1$ and $\Phi_2$ have quantum number $(\textbf{1},\textbf{3})_{-\frac{1}{3}}$.
The gauge kinetic term of $\Phi_1$ and $\Phi_2$ can thus be written as
\begin{equation}
\mL_{gk}=(D_\mu\Phi_1)^\dagger(D^\mu\Phi_1)+
(D_\mu\Phi_2)^\dagger(D^\mu\Phi_2)
\end{equation}
in which the covariant derivative can be expressed as
\begin{equation}
D_\mu=\partial_\mu-igA_\mu^a T^a+ig_xQ_xB_\mu^x,\quad
g_x=\frac{gt_W}{\sqrt{1-t_W^2/3}}
\end{equation}
In the above equation, $A_\mu^a$ and $B_\mu^x$ denote $SU(3)_L$
and $U(1)_X$ gauge fields, respectively. $g$ and $g_x$ denote the coupling
constants of $SU(3)_L$ and $U(1)_X$ gauge groups, respectively.
It is convenient to trade $g_x$ for $t_W\equiv\tan\theta_W$
for future derivation. $T^a=\frac{\lambda^a}{2}$ where
$\lambda^a,a=1,...,8$ denote the Gell-Mann matrices. For $\Phi_1,\Phi_2$,
$Q_x=-\frac{1}{3}$. Following ref.~\cite{delAguila:2011wk}, we
parametrize the $SU(3)_L$ gauge bosons as
\begin{align}
A_\mu^a T^a=\frac{A_\mu^3}{2}
\begin{pmatrix}
1 & 0 & 0 \\
0 & -1 & 0 \\
0 & 0 & 0
\end{pmatrix}
+\frac{A_\mu^8}{2\sqrt{3}}
\begin{pmatrix}
1 & 0 & 0 \\
0 & 1 & 0 \\
0 & 0 & -2
\end{pmatrix}
+\frac{1}{\sqrt{2}}
\begin{pmatrix}
0 & W_\mu^+ & Y_\mu^0 \\
W_\mu^- & 0 & X_\mu^- \\
Y_\mu^{0\dagger} & X_\mu^+ & 0
\end{pmatrix}
\end{align}
with the \textit{first-order} neutral gauge boson mixing relation ($c_W\equiv\cos\theta_W,s_W\equiv\sin\theta_W$)
\begin{align}
\begin{pmatrix}
A^3 \\ A^8 \\ B_x
\end{pmatrix}
=
\begin{pmatrix}
0 & c_W & -s_W \\
\sqrt{1-\frac{t_W^2}{3}} & \frac{s_W t_W}{\sqrt{3}} & \frac{s_W}{\sqrt{3}} \\
-\frac{t_W}{\sqrt{3}} & s_W\sqrt{1-\frac{t_W^2}{3}} & c_W\sqrt{1-\frac{t_W^2}{3}}
\end{pmatrix}
\begin{pmatrix}
Z' \\ Z \\ A
\end{pmatrix}
\end{align}

We note in passing that in the presence
of vacuum misalignment (i.e. $v\neq 0$), generally speaking
$\eta,\zeta,\chi,\omega$ will not be canonically-normalized.
Also, there could exist ``unexpected'' vector-scalar
mixing terms such as $Z_\mu\partial^\mu\eta$ at tree
level. This kind of situation is not uncommon for models
which invoke gauged nonlinear sigma fields. Even if $\zeta,\chi,
\omega$ can be rotated away by gauge transformations,
terms like $Z_\mu\partial^\mu\eta$ certainly cannot be
eliminated by a naive gauge rotation. A systematic procedure
for diagonalizing such a vector-scalar system in gauge theories,
including the elimination of ``unexpected'' vector-scalar
two-point mixing via an appropriate gauge-fixing, is outlined
in ref.~\cite{He:2017jjx} and applied to the SLH. The implication
is that a further transformation among the $\eta,\zeta,\chi,\omega$
has to be made to derive the correct masses and couplings
related to these particles. For the main purpose of the present
paper, this subtlety will only lead to a $\ord(v^2/f^2)$-suppressed
correction to the derived $\eta$ mass.

We now turn to the Yukawa Lagrangian. Since the electroweak
gauge group is now $SU(3)_L\times U(1)_X$, new fermions need
to be introduced to furnish complete representations of
the gauge group. This can be done elegantly in an anomaly-free
manner~\cite{Kong:2003vm,Schmaltz:2004de,Kong:2004cv} which we adopt here.
In the lepton Yukawa sector, the SM left-handed lepton doublets
are enlarged to $SU(3)_L$ triplets $L_m=(\nu_L,\ell_L,iN_L)_m^T$
with $Q_x=-\frac{1}{3}$ ($m=1,2,3$ is the family index). There
are also right-handed singlet lepton fields $\ell_{Rm}$ with $Q_x=-1$
and $N_{Rm}$ with $Q_x=0$. The lepton Yukawa Lagrangian can
be written as~\cite{delAguila:2011wk}
\begin{equation}
\mL_{LY}=i\lambda_N^m\bar{N}_{Rm}\Phi_2^\dagger L_m
+\frac{i\lambda_\ell^{mn}}{\Lambda}\bar{\ell}_{Rm}\epsilon_{ijk}\Phi_1^i\Phi_2^j L_n^k
+\text{h.c.}
\label{eq:lly}
\end{equation}
The charged leptons $e,\mu,\tau$ pick up their masses through
the dimension-five operators in $\mL_{LY}$, in which an energy scale
$\Lambda$ is introduced to make the $3\times 3$ mass matrix
$\lambda_\ell$ dimensionless. The dimension-four operators in
$\mL_{LY}$ makes the new leptons $N_R$'s massive. It should be
noted that light neutrinos $\nu_L$'s remain massless
with $\mL_{LY}$, although their masses can be straightforwardly included
by adding $(\Phi_2^\dagger L)^2$ operators.

The anomaly-free requirement leads to the following
quark-field content
\begin{align}
Q_1 & =(d_L,-u_L,iD_L)^T, \quad d_R,\quad u_R,\quad D_R \\
Q_2 & =(s_L,-c_L,iS_L)^T, \quad s_R,\quad c_R,\quad S_R \\
Q_3 & =(t_L,b_L,iT_L)^T, \quad t_R,\quad b_R,\quad T_R
\end{align}
Here $Q_1,Q_2$ transform under $\bar{\textbf{3}}$ representation
of $SU(3)_L$ with $Q_x=0$. $Q_3$ transforms under $\textbf{3}$
representation of $SU(3)_L$ with $Q_x=\frac{1}{3}$. The right-handed
quark fields are all $SU(3)_L$ singlets with various $U(1)_X$ charges.
More specifically, $u_R,c_R,t_R,T_R$ carry $Q_x=\frac{2}{3}$ while
$d_R,s_R,b_R,D_R,S_R$ carry $Q_x=-\frac{1}{3}$. The quark
Yukawa Lagrangian can be written as~\cite{delAguila:2011wk}
\begin{align}
\mL_{QY} & = i\lambda_1^t\bar{u}_{R3}^1\Phi_1^\dagger Q_3
+i\lambda_2^t\bar{u}_{R3}^2\Phi_2^\dagger Q_3
+i\frac{\lambda_b^m}{\Lambda}\bar{d}_{Rm}\epsilon_{ijk}\Phi_1^i\Phi_2^jQ_3^k \nonumber \\
& +i\lambda_1^{dn}\bar{d}_{Rn}^1Q_n^T\Phi_1
+i\lambda_2^{dn}\bar{d}_{Rn}^2Q_n^T\Phi_2
+i\frac{\lambda_u^{mn}}{\Lambda}\bar{u}_{Rm}\epsilon_{ijk}\Phi_1^{*i}\Phi_2^{*j}Q_n^k
+\text{h.c.}
\label{eq:lqy}
\end{align}
In the above equation, $n=1,2$ is the family index for
the first two generations of quark triplets. $d_{Rm}$ runs over
$(d_R,s_R,b_R,D_R,S_R)$ and $u_{Rm}$ runs over
$(u_R,c_R,t_R,T_R)$. $u_{R3}^1,u_{R3}^2$ are linear combinations
of $t_R$ and $T_R$. $d_{Rn}^1,d_{Rn}^2$ are linear combinations
of $d_R$ and $D_R$ for $n=1$ and of $s_R$ and $S_R$ for $n=2$.

The CSB mechanism in the SLH deserves comment. In the bosonic
sector, $\mL_{gk}$ automatically realizes CSB, while in the
fermionic sector, we have deliberately chosen the dimension-four
operators in $\mL_{LY},\mL_{QY}$ to ensure CSB. Especially,
in Eq.~\eqref{eq:lly} we do not write down a
$\bar{N}_{R}\Phi_1^\dagger L$ term which is allowed by gauge
symmetry but formally violates CSB when $\bar{N}_{R}\Phi_2^\dagger L$
is also present. In Eq.~\eqref{eq:lqy}, the crucial ingredient
for scalar potential analysis is the top sector Lagrangian
\begin{equation}
\mL_{tY} = i\lambda_1^t\bar{u}_{R3}^1\Phi_1^\dagger Q_3
+i\lambda_2^t\bar{u}_{R3}^2\Phi_2^\dagger Q_3 \subset \mL_{QY}
\end{equation}
in which the CSB is manifest. The dimension-five operators
in Eq.~\eqref{eq:lly} and Eq.~\eqref{eq:lqy} actually
violate CSB. Nevertheless, these sources of violation
are proportional to light fermion Yukawa and their
effect on scalar potential analysis can be safely neglected.

If in Eq.~\eqref{eq:lly} the $i\lambda_N^m\bar{N}_{Rm}\Phi_2^\dagger L_m$
term is neglected at the moment, then we could restrict
the range of $t_\beta$ to be $t_\beta\geq 1$ without loss
of generality. This is because in this case we are always free to
label the scalar triplet with smaller vacuum expectation value
as $\Phi_1$. However, when we require the very presence of
$i\lambda_N^m\bar{N}_{Rm}\Phi_2^\dagger L_m$ term in
Eq.~\eqref{eq:lly}, this labelling redundancy does not
hold any more, and we need to consider both
$t_\beta\geq 1$ and $t_\beta<1$. Nevertheless, for the
analysis in the present paper, it is found that
the labelling redundancy $t_\beta\leftrightarrow\frac{1}{t_\beta}$
still holds to a good approximation since the correction
only comes in at $\ord(v^2/f^2)$ in the input parameter.
Therefore, in the rest of this paper we will still
present the results by focusing on the $t_\beta\geq 1$ case.
\section{Hidden Mass Relation from Scalar Potential Analysis}
\label{sec:hmr}
\subsection{Scalar Potential in the SLH}

Up to now, we have not described the scalar potential
in the SLH yet. In Section~\ref{sec:intro}, we mentioned
two approaches commonly adopted in the literature.
In one of them, the relevant model parameters are treated
effectively as free parameters and the predictivity
is lost to a large extent. Therefore, let us scrutinize
the other approach (used in ref.~\cite{Schmaltz:2004de,Reuter:2012sd})
to see whether there is room for improvement.

In the approach adopted by ref.~\cite{Schmaltz:2004de,Reuter:2012sd},
the tree-level scalar potential is assumed to vanish
except for the $\mu$ term
\begin{equation}
\mL_\mu=\mu^2(\Phi_1^\dagger\Phi_2+\text{h.c.})
\end{equation}
The Yukawa and gauge interactions then generate
a potential at one-loop level, triggering EWSB. At one-loop,
the effective-potential calculation contains
logarithmic UV divergence due to fermion and gauge boson
loops. In ref.~\cite{Schmaltz:2004de,Reuter:2012sd},
this logarithmic divergence is treated by imposing
a momentum cutoff $\Lambda$, which is taken to be
$4\pi f_1$ in ref.~\cite{Schmaltz:2004de} and $4\pi f$
in ref.~\cite{Reuter:2012sd}.

From a CEFT point of view, the appearance of
UV divergence in the calculation signals the need for
renormalization. A similar but simpler example is scalar
quantum electrodynamics (QED). If we do not write down
a scalar quartic term at tree level, then when doing
one-loop calculation of scalar scattering processes we
will still encounter UV divergence which needs to be
absorbed by adding counterterms. From the viewpoint
of renormalization theory, a consistent approach is
to introduce in the bare Lagrangian a scalar quartic
term. Calculation at one-loop order can then be done
via renormalized perturbation theory, in which the bare
Lagrangian is split into the renormalized part and
the counterterm part. The UV divergences encountered
in loop calculations can then be absorbed by counterterms
in appropriate renormalization schemes. The renormalization
procedure will usually introduce an unphysical scale
(renormalization scale) into calculation. Requiring physical
quantities to be independent of this unphysical scale
leads to the notion of running couplings as a consequence
of solving the relevant Callan-Symanzik equation.

Nevertheless, in the literature, a crude momentum cutoff
$\Lambda$ is imposed, and there is
an ambiguity concerning the interpretation of $\Lambda$.
In the Appendix of ref.~\cite{Schmaltz:2004de}, as can be inferred
from the constants appearing in the expression of Coleman-Weinberg
(CW) potential and the text, the CW potential expression
corresponds to $\overline{\text{DR}}$ scheme~\cite{Siegel:1979wq,Quiros:1999jp}
rather than a sharp momentum cutoff regularization without renormalization.
In ref.~\cite{Casas:2005ev}, $\overline{\text{MS}}$ scheme is
used and $\Lambda=4\pi f_1$ is interpreted as the renormalization
scale. However, if $\Lambda$ is regarded as the renormalization scale,
then the choice of its value should be arbitrary in principle since
physical predictions should be renormalization group invariant.
This seems to contradict the fact that the EWSB predictions
in the previous literature indeed rely on say, setting $\Lambda=4\pi f_1$.
As will be discussed in Section~\ref{sec:nislh}, the contradiction disappears
only if we are forced to accept the assumption of vanishing
contribution to the scalar potential from the physics at the cutoff.
In spite of this, interpreting $\Lambda$ as a renormalization scale
is still in conflict with the fact that in ref.~\cite{Schmaltz:2004de,
Casas:2005ev} (and many other papers) a $\Lambda^2$ term is sometimes
retained to discuss cancellation of quadratic divergence or related issues.

In the present paper we opt for an approach consistent with
the spirit of CEFT. We first write down the bare scalar potential
$V_B$ as follows
\begin{equation}
V_B=-\mu_B^2(\Phi_{1B}^\dagger\Phi_{2B}+\Phi_{2B}^\dagger\Phi_{1B})
+\lambda_B |\Phi_{1B}^\dagger\Phi_{2B}|^2
\label{eq:vb}
\end{equation}
The subscript ``$B$'' denotes bare quantities. In $V_B$ we retain
operators up to dimension-four and assume the effects of higher-dimensional
operators can be neglected (which is \textit{not} an inconsistent
power-counting assumption). We note that both operators in
Eq.~\eqref{eq:vb} violate CSB. However, this violation is
not really harmful. For the $\mu$ term, if we neglect the
small light fermion Yukawa, it softly breaks the global
$U(1)$ symmetry in which two scalar triplets undergo
opposite phase rotations. (The pseudo-axion $\eta$ actually
corresponds to the pseudo-Goldstone of this spontaneously
broken $U(1)$.) The scalar quartic coupling $\lambda_B$
is a dimensionless parameter and will not induce a
serious fine-tuning problem unless its renormalized
counterpart takes some extreme value (this issue will
be discussed in more detail in Section~\ref{sec:nislh}).
We emphasize that the inclusion of the scalar
quartic operator $|\Phi_{1B}^\dagger\Phi_{2B}|^2$
is required by renormalization. It provides the necessary
counterterm to absorb the UV divergence encountered
in the calculation of radiative correction to the
scalar effective potential. A dimension-four operator
$(\Phi_{1B}^\dagger\Phi_{2B})^2+\text{h.c.}$ is also
allowed by gauge symmetry, however it would formally cause
a hard breaking of the global $U(1)$ symmetry
that protects the $\mu$ term. Therefore we do not
include it in $V_B$.

We now move on to the calculation of scalar effective potential
via renormalized perturbation theory. The tree-level
effective potential is now written as
\begin{equation}
V_\tree=-\mu^2(\Phi_1^\dagger\Phi_2+\Phi_2^\dagger\Phi_1)
+\lambda_R |\Phi_1^\dagger\Phi_2|^2
\end{equation}
In the above equation all quantities (couplings, fields)
are renormalized ones, with $\Phi_1,\Phi_2$ assuming
the parametrization used in Eq.~\eqref{eq:phi1} and
Eq.~\eqref{eq:phi2}. At one-loop level, we take into
account the contribution from gauge interaction
and top sector Yukawa, and express the scalar potential
at small field value (i.e. $\hat{h}\ll f$) as
\begin{equation}
V_\onel=V_\onel^\ts+V_\onel^{\tns}
\end{equation}
in which the $SU(3)$-symmetric part $V_\onel^\ts$
and $SU(3)$-nonsymmetric part $V_\onel^{\tns}$
are respectively given by
\begin{align}
V_\onel^\ts & = \bar{\lambda}|\Phi_1^\dagger\Phi_2|^2 \label{eq:blambda} \\
V_\onel^{\tns} & = \Delta(\hat{h})\hat{h}^4
\end{align}
Here the coefficient of the $\hat{h}^4$ term is
written as $\Delta(\hat{h})$, indicating that it is field-dependent.
We therefore call $\Delta(\hat{h})$ a \textit{field form factor},
emphasizing it is not a field-independent constant.
Combining the tree-level and one-loop contributions,
the scalar effective potential $V$ (defined as the sum of
$V_\tree$ and $V_\onel$) is given by
\begin{equation}
V=-\mu^2(\Phi_1^\dagger\Phi_2+\Phi_2^\dagger\Phi_1)
+\lambda |\Phi_1^\dagger\Phi_2|^2
+\Delta(\hat{h})\hat{h}^4
\label{eq:vrc}
\end{equation}
in which
\begin{equation}
\lambda\equiv\lambda_R+\bar{\lambda}
\end{equation}
In our treatment the counterterm contribution is included
in $V_\onel$, therefore $\mu^2,\lambda,\Delta(\hat{h})$ in
Eq.~\eqref{eq:vrc} are all finite quantities with no
dependence on the regularization cutoff.

Taking into account gauge boson and top sector Yukawa
contributions, $\bar{\lambda}$ and $\Delta(\hat{h})$
are computed to be (Landau gauge and $\overline{\text{MS}}$
scheme are adopted)
\begin{align}
\bar{\lambda} & = -\frac{3}{8\pi^2}\Bigg[\lambda_t^2
\frac{M_T^2}{f^2}\left(\ln\frac{M_T^2}{\mu_R^2}-1\right)
-\frac{1}{4}g^2\frac{M_X^2}{f^2}\left(\ln\frac{M_X^2}{\mu_R^2}-\frac{1}{3}\right) \nonumber \\
& -\frac{1}{8}g^2 (1+t_W^2)\frac{M_{Z'}^2}{f^2}\left(\ln\frac{M_{Z'}^2}{\mu_R^2}-\frac{1}{3}\right)
\Bigg]
\label{eq:bl}
\end{align}
\begin{align}
\Delta(\hat{h}) & =\frac{3}{16\pi^2}\Bigg\{\lambda_t^4
\left[\ln\frac{M_T^2}{m_t^2(\hat{h})}-\frac{1}{2}\right]
-\frac{1}{8}g^4\left[\ln\frac{M_X^2}{m_W^2(\hat{h})}-\frac{1}{2}\right] \nonumber \\
& -\frac{1}{16}g^4(1+t_W^2)^2\left[\ln\frac{M_{Z'}^2}{m_Z^2(\hat{h})}-\frac{1}{2}\right]
\Bigg\}
\label{eq:dh}
\end{align}
In the above equations, $\mu_R$ is the renormalization scale
in the $\overline{\text{MS}}$ scheme. We deliberately avoid the use of
$\Lambda$ here to prevent any interpretational ambiguity. $\lambda_t$ is defined as
\begin{equation}
\lambda_t\equiv\frac{\lambda_1^t\lambda_2^t}
{\sqrt{\lambda_1^{t2} c_\beta^2+\lambda_2^{t2}s_\beta^2}}
\label{eq:lt1}
\end{equation}
where $\lambda_1^t,\lambda_2^t$ are the two
Yukawa couplings in the top sector, introduced in
Eq.~\eqref{eq:lqy}. $M_T^2,M_X^2,M_{Z'}^2$ are defined as
\begin{align}
M_T^2 & \equiv(\lambda_1^{t2} c_\beta^2+\lambda_2^{t2}s_\beta^2)f^2 \\
M_X^2 & \equiv\frac{1}{2}g^2 f^2 \\
M_{Z'}^2 & \equiv\frac{2}{3-t_W^2}g^2 f^2 \label{eq:zpmass}
\end{align}
They are related to physical mass squared of the relevant particles
as follows
\begin{align}
M_T^2 & =m_T^2+m_t^2 \label{eq:MT} \\
M_X^2 & =m_X^2+m_W^2 \\
M_{Z'}^2 & =m_{Z'}^2+m_Z^2
\end{align}
in which $m_T,m_t$ denote the physical mass of the heavy top $T$
and the top quark $t$, $m_X,m_W$ denote the physical mass of
the $X$ boson and $W$ boson, $m_{Z'},m_Z$ denote the physical mass
of the $Z'$ boson and $Z$ boson, respectively. We distinguish between
$M_T$ and $m_T$, $M_X$ and $m_X$, $M_{Z'}$ and $m_{Z'}$,
although the numerical differences are very small. $m_t^2(\hat{h}),
m_W^2(\hat{h}),m_Z^2(\hat{h})$ are field-dependent mass squared,
which we use the following leading order expression in the field
form factor
\begin{align}
m_t^2(\hat{h}) & =\lambda_t^2\hat{h}^2 \\
m_W^2(\hat{h}) & =\frac{1}{2}g^2\hat{h}^2 \\
m_Z^2(\hat{h}) & =\frac{1}{2}g^2(1+t_W^2)\hat{h}^2
\end{align}
We will see that retaining the field-dependence in these
expressions is important for the quantitative study of $m_\eta-m_T$
correlation.

We note that in our calculation of $V$, the gauge boson and fermionic
contributions are considered at one-loop order, while the contribution
from the scalar sector itself is only considered to tree level. This is
consistent since the leading contribution of gauge boson and fermion fields to
the scalar effective potential arise at one-loop order while the leading
contribution of scalar fields to the scalar effective potential arises already
at tree level in our treatment. As long as perturbation theory is valid,
the scalar one-loop contribution should always be small compared to the
scalar tree level contribution which we already take into account.

\subsection{Analysis of the Scalar Effective Potential}

Having obtained the scalar effective potential, Eq.~\eqref{eq:vrc},
with the expression of $\bar{\lambda}$ and $\Delta(\hat{h})$ given in
Eq.~\eqref{eq:bl} and Eq.~\eqref{eq:dh}, we now begin to analyze its physical
implications. It is helpful to first pin down the dimension of parameter
space that we are dealing with. For the purpose of scalar potential analysis,
if we consider $m_t,g$ and $t_W$ as known and fixed, then before fixing $v$
and the CP-even Higgs mass (denoted as $m_h$), we have five adjustable
parameters that enter into $V$. For instance, we may choose the five
parameters to be $f,t_\beta,M_T,\mu,\lambda$. Once these five parameters
are given, other quantities like $v,m_h$ and $\eta$ mass $m_\eta$ can be
derived. Alternatively, we can utilize the measured value of electroweak
vacuum expectation value and CP-even Higgs mass to eliminate two of
the five initial parameters (say, $\mu$ and $\lambda$), leaving
the remaining three $f,t_\beta,M_T$ as independent parameters to
characterize the parameter space. The $\eta$ mass, which is associated with
a second derivative of $V$ at its local minimum, can be determined from
$f,t_\beta,M_T$. This is exactly the hidden mass relation that we
wish to point out in the present paper ($M_T$ is related to $m_T$ via
Eq.~\eqref{eq:MT}). This relation can be viewed as a $m_\eta-m_T$ mass
relation given $f$ and $t_\beta$, and serve as a crucial test of
the SLH mechanism. This relation can also be viewed as a $m_\eta-m_h$
mass relation. When we consider only tree-level scalar effective
potential, an $m_\eta-m_h$ mass relation can be obtained which is
valid for any value of the Lagrangian parameters. When one-loop
gauge boson and fermion contributions are taken into account,
the correction to $m_\eta-m_h$ mass relation is then automatically
finite and becomes a calculable prediction of the symmetry structure
of the theory. In this sense, this hidden mass relation is
a \textit{zeroth-order natural relation}~\cite{Peskin:1995ev}, which
we now begin to derive.

A convenient starting point is the scalar effective potential $V$ in
Eq.~\eqref{eq:vrc} at small field value. With the parametrization
Eq.~\eqref{eq:phi1} and Eq.~\eqref{eq:phi2}, $\Phi_1^\dagger\Phi_2$
can be expressed in terms of $\hat{h}$ and $\eta$ exactly as
follows~\cite{Cheung:2006nk}
\begin{equation}
\Phi_1^\dagger\Phi_2=f^2s_\beta c_\beta\exp\left(
-\frac{i}{\sqrt{2}fs_\beta c_\beta}\eta\right)\cos\left(
\frac{1}{fs_\beta c_\beta}\hat{h}\right)
\end{equation}
Making use of this expression we may express $V$ in
Eq.~\eqref{eq:vrc} as a function of two real variables
$\hat{h}$ and $\eta$
\begin{align}
V & =-2\mu^2 f^2 s_\beta c_\beta\cos\left(
\frac{1}{\sqrt{2}fs_\beta c_\beta}\eta\right)\cos\left(
\frac{1}{fs_\beta c_\beta}\hat{h}\right)
+\lambda f^4 s_\beta^2 c_\beta^2 \cos^2\left(
\frac{1}{fs_\beta c_\beta}\hat{h}\right) \nonumber \\
& +\Delta\hat{h}^4
\end{align}
Here and in the following, we simply use $\Delta$ to
represent the field form factor $\Delta(\hat{h})$,
keeping in mind its field-dependence. The electroweak vacuum
should correspond to a stationary point of $V$, satisfying
$\frac{\partial V}{\partial\eta}=\frac{\partial V}{\partial\hat{h}}
=0$, i.e.
\begin{align}
\sqrt{2}\mu^2 f\sin\left(
\frac{1}{\sqrt{2}fs_\beta c_\beta}\eta\right)\cos\left(
\frac{1}{fs_\beta c_\beta}\hat{h}\right)=0
\label{eq:vs1}
\end{align}
\begin{align}
& 2\mu^2 f\cos\left(
\frac{1}{\sqrt{2}fs_\beta c_\beta}\eta\right)\sin\left(
\frac{1}{fs_\beta c_\beta}\hat{h}\right)-\lambda f^3 s_\beta
c_\beta\sin\left(\frac{2}{fs_\beta c_\beta}\hat{h}\right)
+4\Delta\hat{h}^3 \nonumber \\
& +\Delta'\hat{h}^4=0
\label{eq:vs2}
\end{align}
in which
\begin{equation}
\Delta'\equiv\frac{d\Delta}{d\hat{h}}
\end{equation}
We are concerned with the case of no spontaneous CP-violation,
i.e. $\eta=0$. In this case, Eq.~\eqref{eq:vs1} is automatically
satisfied, and Eq.~\eqref{eq:vs2} becomes
\begin{equation}
2\mu^2 f\sin\left(
\frac{1}{fs_\beta c_\beta}\hat{h}\right)-\lambda f^3 s_\beta
c_\beta\sin\left(\frac{2}{fs_\beta c_\beta}\hat{h}\right)
+4\Delta\hat{h}^3+\Delta'\hat{h}^4=0
\label{eq:spc}
\end{equation}
Suppose $P_0=(0,\hat{h}_0)$ corresponds to the $(\eta,\hat{h})$
field configuration of the electroweak vacuum and therefore
$\hat{h}_0$ is a solution of Eq.~\eqref{eq:spc}. According to
Eq.~\eqref{eq:theta} $\hat{h}_0$ is related to $v$ by
\begin{equation}
\hat{h}_0=\frac{v}{\sqrt{2}}
\end{equation}
The elements of the Hessian matrix of $V$ at point $P_0$ are computed
to be
\begin{align}
\frac{\partial^2 V}{\partial\eta^2}\Bigg|_{P_0}
& =\frac{\mu^2}{s_\beta c_\beta}\cos\left(
\frac{1}{fs_\beta c_\beta}\hat{h}_0\right) \\
\frac{\partial^2 V}{\partial\eta\partial\hat{h}}\Bigg|_{P_0}
& =0 \\
\frac{\partial^2 V}{\partial\hat{h}^2}\Bigg|_{P_0}
& =\frac{2\mu^2}{s_\beta c_\beta}\cos\left(
\frac{1}{fs_\beta c_\beta}\hat{h}_0\right)
-2\lambda f^2\cos\left(\frac{2}{fs_\beta c_\beta}\hat{h}_0\right)
+12\Delta_0\hat{h}_0^2+8\Delta_0'\hat{h}_0^3 \nonumber \\
& +\Delta_0''\hat{h}_0^4
\end{align}
in which
\begin{equation}
\Delta_0\equiv\Delta(\hat{h}_0),\quad
\Delta_0'\equiv\frac{d\Delta}{d\hat{h}}\Bigg|_{\hat{h}_0},\quad
\Delta_0''\equiv\frac{d^2\Delta}{d\hat{h}^2}\Bigg|_{\hat{h}_0}
\end{equation}
Since the off-diagonal entry of the Hessian matrix is zero,
we could read out the pseudo-axion mass $m_\eta$ and Higgs mass
$m_h$ directly from the above equations. The only subtlety is that
the $\eta$ field introduced
in Eq.~\eqref{eq:theta} is not canonically-normalized. It
is related to the canonically-normalized mass eigenstate
field $\eta^m$ by a simple rescaling relation~\cite{He:2017jjx}
\begin{equation}
\eta=\eta^m\sec\left(\frac{\hat{h}_0}{fs_\beta c_\beta}\right)
\end{equation}
With this in mind, the pseudo-axion and Higgs mass squared
are found to be
\begin{align}
m_\eta^2=\frac{\mu^2}{s_\beta c_\beta}
\sec\left(\frac{\hat{h}_0}{fs_\beta c_\beta}\right)
\end{align}
\begin{align}
m_h^2=\frac{\mu^2}{s_\beta c_\beta}\cos\left(
\frac{1}{fs_\beta c_\beta}\hat{h}_0\right)
-\lambda f^2\cos\left(\frac{2}{fs_\beta c_\beta}\hat{h}_0\right)
+6\Delta_0\hat{h}_0^2+4\Delta_0'\hat{h}_0^3
+\frac{1}{2}\Delta_0''\hat{h}_0^4
\label{eq:mh2}
\end{align}
We could obtain an $m_\eta-m_h$ mass relation by
eliminating the $\lambda$ in Eq.~\eqref{eq:mh2}
using the stationary point condition Eq.~\eqref{eq:spc},
and the result is (using $v=\sqrt{2}\hat{h}_0$)
\begin{align}
m_\eta^2 & =\Bigg\{m_h^2-v^2\Delta_0\left[3-\frac{\sqrt{2}v}{fs_\beta c_\beta}
\cot\left(\frac{\sqrt{2}v}{fs_\beta c_\beta}\right)\right] \nonumber \\
& -\frac{1}{4}v^3\Delta_0'\left[4\sqrt{2}-\frac{v}{fs_\beta c_\beta}
\cot\left(\frac{\sqrt{2}v}{fs_\beta c_\beta}\right)\right]
-\frac{1}{8}v^4\Delta_0''\Bigg\}
\csc^2\left(\frac{v}{\sqrt{2}fs_\beta c_\beta}\right)
\label{eq:mr1}
\end{align}
From Eq.~\eqref{eq:dh} we may easily obtain the following
expressions for $\Delta_0,\Delta_0',\Delta_0''$
\begin{align}
\Delta_0 & =\frac{3}{16\pi^2}\Bigg[\lambda_t^4\left(
\ln\frac{M_T^2}{m_t^2}-\frac{1}{2}\right)-\frac{g^4}{8}
\left(\ln\frac{M_X^2}{m_W^2}-\frac{1}{2}\right) \nonumber \\
& -\frac{g^4}{16}(1+t_W^2)^2\left(\ln\frac{M_{Z'}^2}{m_Z^2}-\frac{1}{2}\right)
\Bigg]
\label{eq:delta0}
\end{align}
\begin{align}
\Delta_0'=-\frac{3\sqrt{2}}{8\pi^2 v}
\left[\lambda_t^4-\frac{g^4}{8}-\frac{g^4}{16}(1+t_W^2)^2\right]
\end{align}
\begin{align}
\Delta_0''=\frac{3}{4\pi^2 v^2}
\left[\lambda_t^4-\frac{g^4}{8}-\frac{g^4}{16}(1+t_W^2)^2\right]
\end{align}
The $m_t^2,m_W^2,m_Z^2$ in the expression of $\Delta_0$ are
field-independent and correspond to the physical mass squared
of the top quark, $W$ and $Z$ bosons, respectively. In the mass
relation Eq.~\eqref{eq:mr1}, the field-dependent
effects of the field form factor $\Delta(\hat{h})$ are
encoded in $\Delta_0',\Delta_0''$.

If we define
\begin{align}
\theta\equiv\frac{v}{\sqrt{2}fs_\beta c_\beta}
\end{align}
\begin{align}
A\equiv\frac{3}{16\pi^2}
\left[\lambda_t^4-\frac{g^4}{8}-\frac{g^4}{16}(1+t_W^2)^2\right]
\label{eq:A}
\end{align}
\begin{align}
\Delta_A\equiv\frac{3}{16\pi^2}\left[\lambda_t^4
\ln\frac{M_T^2}{m_t^2}-\frac{g^4}{8}\ln\frac{M_X^2}{m_W^2}
-\frac{g^4}{16}(1+t_W^2)^2\ln\frac{M_{Z'}^2}{m_Z^2}\right]
\end{align}
then the $m_\eta-m_h$ mass relation can be written as
\begin{align}
m_\eta^2=[m_h^2-v^2\Delta_A(3-2\theta t_{2\theta}^{-1})
+v^2 A(5-2\theta t_{2\theta}^{-1})]s_\theta^{-2}
\label{eq:mr2}
\end{align}
Here $t_{2\theta}^{-1}\equiv\frac{1}{\tan(2\theta)},
s_\theta^{-2}\equiv\frac{1}{\sin^2\theta}$. Eq.~\eqref{eq:mr2}
is the central result of this paper. If we set $\Delta_A=A=0$,
then Eq.~\eqref{eq:mr2} would yield the corresponding prediction
from considering only tree-level scalar potential. The correction
to the tree level prediction as exhibited by Eq.~\eqref{eq:mr2},
is obviously finite and does not depend on the renormalization
scale manifestly, consistent with the expectation for
a zeroth-order natural relation.

From Eq.~\eqref{eq:mr2} we may deduce that $m_\eta$ can be predicted
once $f,t_\beta$ and $m_T$ are known. This prediction can be obtained
as long as the SLH can be treated as a self-contained EFT in its
domain of validity. It can be tested in a quantitative manner
without ad hoc assumptions about UV physics contribution. We see
that $m_\eta$ and $m_T$ are anti-correlated, i.e. with a heavier
$m_T$ we will obtain a lighter $m_\eta$. The electroweak vacuum
is supposed to be a local minimum of the scalar effective potential,
and therefore we require $m_\eta^2\geq 0$, which is equivalent to
\begin{align}
m_h^2-v^2\Delta_A(3-2\theta t_{2\theta}^{-1})
+v^2 A(5-2\theta t_{2\theta}^{-1})\geq 0
\label{eq:poseta}
\end{align}
This condition sets an upper bound on $m_T$ when $f,t_\beta$ are given.
The parameter $M_T$ also has a lower bound for fixed $f,t_\beta$
~\cite{Han:2005ru}
\begin{equation}
M_T\geq \sqrt{2}\frac{m_t}{v}fs_{2\beta}\approx fs_{2\beta}
\label{eq:MTmin}
\end{equation}
where $s_{2\beta}\equiv\sin(2\beta)$. This bound comes from the
definition of $M_T$ in Eq.~\eqref{eq:MT} and the requirement of
the top Yukawa $\lambda_t$ in Eq.~\eqref{eq:lt1} to yield the
correct top quark mass.

\subsection{Unitarity Constraint}

Due to the nonlinearly-realized scalar sector, the SLH can at best
be seen as an EFT valid up to some cutoff scale. Apart from NDA
consideration which yields a cutoff of $4\pi fc_\beta$, we may also
consider the bound from perturbative unitarity, which is
usually expected to yield a more stringent constraint compared to NDA
~\cite{Chang:2003vs}.

To derive the perturbative unitarity constraint for the SLH, we
adopt the methodology of ref.~\cite{Chang:2003vs}. To simplify the
problem, we only consider the kinetic terms of the nonlinear sigma
fields, in the limit of vanishing EWSB. In the SLH we have two copies
of $SU(3)\rightarrow SU(2)$ global symmetry breaking, realized by
$\Phi_1$ and $\Phi_2$ respectively, and the only relevant difference
between them is the symmetry breaking scale, $fc_\beta$ for $\Phi_1$
and $fs_\beta$ for $\Phi_2$. In our setting $\Phi_1$ is chosen to be
the one with a lower symmetry breaking scale (i.e. $t_\beta\geq 1$),
and therefore we only consider the perturbative unitarity constraint
from analysis of kinetic terms of $\Phi_1$.

For the perturbative unitarity analysis, it is convenient to
parameterize $\Phi_1$ as ($f_1\equiv fc_\beta$)
\begin{align}
\Phi_1=\exp\left(\frac{i\Theta_1}{f_1}\right)
\begin{pmatrix}
0 \\ 0 \\ f_1
\end{pmatrix}
\end{align}
in which
\begin{align}
\Theta_1=\frac{1}{\sqrt{2}}
\begin{pmatrix}
\frac{1}{2}\pi_8 & 0 & \pi_4-i\pi_5 \\
0 & \frac{1}{2}\pi_8 & \pi_6-i\pi_7 \\
\pi_4+i\pi_5 & \pi_6+i\pi_7 & -\pi_8
\end{pmatrix}
\end{align}
The kinetic Lagrangian of $\Phi_1$ can now be written as
\begin{align}
\mL_{k1} & =(\partial_\mu\Phi_1)^\dagger\partial^\mu\Phi_1 \nonumber \\
& =\frac{1}{2}\sum_a\partial_\mu\pi_a\partial^\mu\pi_a
+\frac{1}{4\sqrt{2}f_1}(\pi_4\partial_\mu\pi_5-\pi_5\partial_\mu\pi_4
+\pi_6\partial_\mu\pi_7-\pi_7\partial_\mu\pi_6)\partial^\mu\pi_8 \nonumber \\
& +\sum_{i<j}\left\{-\frac{1}{12f_1^2}[\pi_i^2(\partial_\mu\pi_j)^2
+\pi_j^2(\partial_\mu\pi_i)^2]+\frac{1}{6f_1^2}(\pi_i\partial_\mu\pi_i)
(\pi_j\partial^\mu\pi_j)\right\} \nonumber \\
& +\sum_i\left\{-\frac{3}{64f_1^2}\pi_8^2(\partial_\mu\pi_i)^2
-\frac{7}{64f_1^2}(\partial_\mu\pi_8)^2\pi_i^2
+\frac{5}{32f_1^2}(\pi_8\partial_\mu\pi_8)(\pi_i\partial^\mu\pi_i)\right\}
\end{align}
In the above equation, the ranges of summation are $a=4,5,6,7,8$ and
$i,j=4,5,6,7$. Considering $\pi_a\pi_a\rightarrow\pi_b\pi_b$ scattering
(with $a,b=4,5,6,7,8$), the 0-th partial wave amplitude matrix is then computed
to be
\begin{align}
\mathcal{A}_0=\frac{s}{64\pi f_1^2}
\begingroup 
\setlength\arraycolsep{4pt}
\begin{pmatrix}
0 & 1 & 1 & 1 & 1 \\
1 & 0 & 1 & 1 & 1 \\
1 & 1 & 0 & 1 & 1 \\
1 & 1 & 1 & 0 & 1 \\
1 & 1 & 1 & 1 & 0
\end{pmatrix}
\endgroup
\end{align}
The eigenvalues of $\mathcal{A}_0$ are
\begin{align}
a_{0j}=\frac{s}{64\pi f_1^2}(4,-1,-1,-1,-1)
\end{align}
Requiring $|\text{Re}(a_{0j})|\leq\frac{1}{2}$, we
are able to obtain the perturbative unitarity constraint
\begin{align}
\sqrt{s}\leq\sqrt{8\pi}f_1
\end{align}
As expected, this turns out to be more stringent than
the NDA bound of $\sqrt{s}\leq4\pi f_1$. In the following analysis,
we require the particle masses that are relevant to our
scalar potential analysis be smaller than the
unitarity cutoff $\sqrt{8\pi}f_1$. Specifically, we require
(recall $f_1\equiv fc_\beta$)
\begin{align}
M_{Z'}\leq\sqrt{8\pi}fc_\beta \label{eq:zpu1} \\
M_T\leq\sqrt{8\pi}fc_\beta \label{eq:mtu}
\end{align}
In the SLH, $Z'$ is heavier than $X,Y$ bosons, therefore we impose
the constraint on $M_{Z'}$. The difference between $M_{Z'}$ and
$m_{Z'}$ and the difference between $M_T$ and $m_T$ are both small,
and in the unitarity constraint we use $M_{Z'}$ and $M_T$ for
simplicity. We can use Eq.~\eqref{eq:zpmass} to find a constraint
on $t_\beta$ from Eq.~\eqref{eq:zpu1}. This constraint when
combined with the assumption of $t_\beta\geq 1$, implies the
following allowed region of $t_\beta$ which we assume hereafter
\begin{align}
1\leq t_\beta\leq\sqrt{\frac{4\pi(3-t_W^2)}{g^2}-1}
\label{eq:tbrange}
\end{align}
The unitarity constraint yields an upper bound on $t_\beta$,
which is not difficult to understand. $t_\beta$ can be viewed as
a measure of the asymmetry between the vacuum expectation values
of $\Phi_1$ and $\Phi_2$. The unitarity constraint depends on
the smaller of the two vacuum expectation values, while the $Z'$
boson mass depends on their quadrature. Therefore the asymmetry
between the two vacuum expectation values cannot be too large.

\subsection{Allowed Region of Parameter Space}

With the results from the scalar potential analysis and unitarity
constraints obtained in this section, we are ready to make plots
characterizing the allowed region of parameter space with an
understanding of its basic features. However, before that we
note there are some technicalities related to input parameter
corrections. More specifically, the parameters $g,v,t_W,\lambda_t$
that appear in our parametrization of the SLH should not be
directly identified with the corresponding quantities in the SM,
which we denote as $g_\SM,v_\SM,t_{W,\SM},\lambda_{t,\SM}$, respectively. Their
relations should be established by producing a common set of
well-measured physical observables. In this paper, we are concerned
with correction due to the SLH, rather than higher order radiative
corrections. Thus we opt to work at tree level but retain the leading
$\ord(\frac{v^2}{f^2})$ corrections. The correspondence turns out
to be
\begin{align}
g & =g_\SM\left(1+\frac{1}{4t_\beta^2}\frac{v_\SM^2}{f^2}\right) \\
v & =v_\SM\left(1+\frac{t_\beta^4-t_\beta^2-2}{12t_\beta^2}\frac{v_\SM^2}{f^2}\right) \\
t_W & =t_{W,\SM}\left(1-\frac{1+t_{W,\SM}^2}{4t_\beta^2}\frac{v_\SM^2}{f^2}\right) \\
\lambda_t & =\lambda_{t,\SM}\left[1+\left(\frac{1}{4s_\beta^2}
-\frac{\lambda_{t,\SM}^2}{4}\frac{f^2}{M_T^2}\right)\frac{v_\SM^2}{f^2}\right]
\end{align}
in which we use the following values for SM quantities
\begin{align}
g_\SM=0.653,\quad v_\SM=246.2\GeV,\quad t_{W,\SM}=0.536,\quad \lambda_{t,\SM}=0.995
\end{align}
The $Z'$ boson in the SLH is subject to the stringent constraint from
the LHC search of a high-mass resonance decaying into dilepton.
Using ATLAS results~\cite{Aaboud:2017buh} we set a crude lower bound on $f$ as~\cite{Mao:2017hpp}
\begin{align}
f>7.5\TeV
\end{align}
With such a stringent constraint the effect of $\ord(\frac{v^2}{f^2})$ corrections
is in fact very small. Nevertheless we take them into account in our numerical analysis
to validate the stability of our results against these corrections.

\begin{figure*}[ht]
\begin{center}
\includegraphics[width=2.5in]{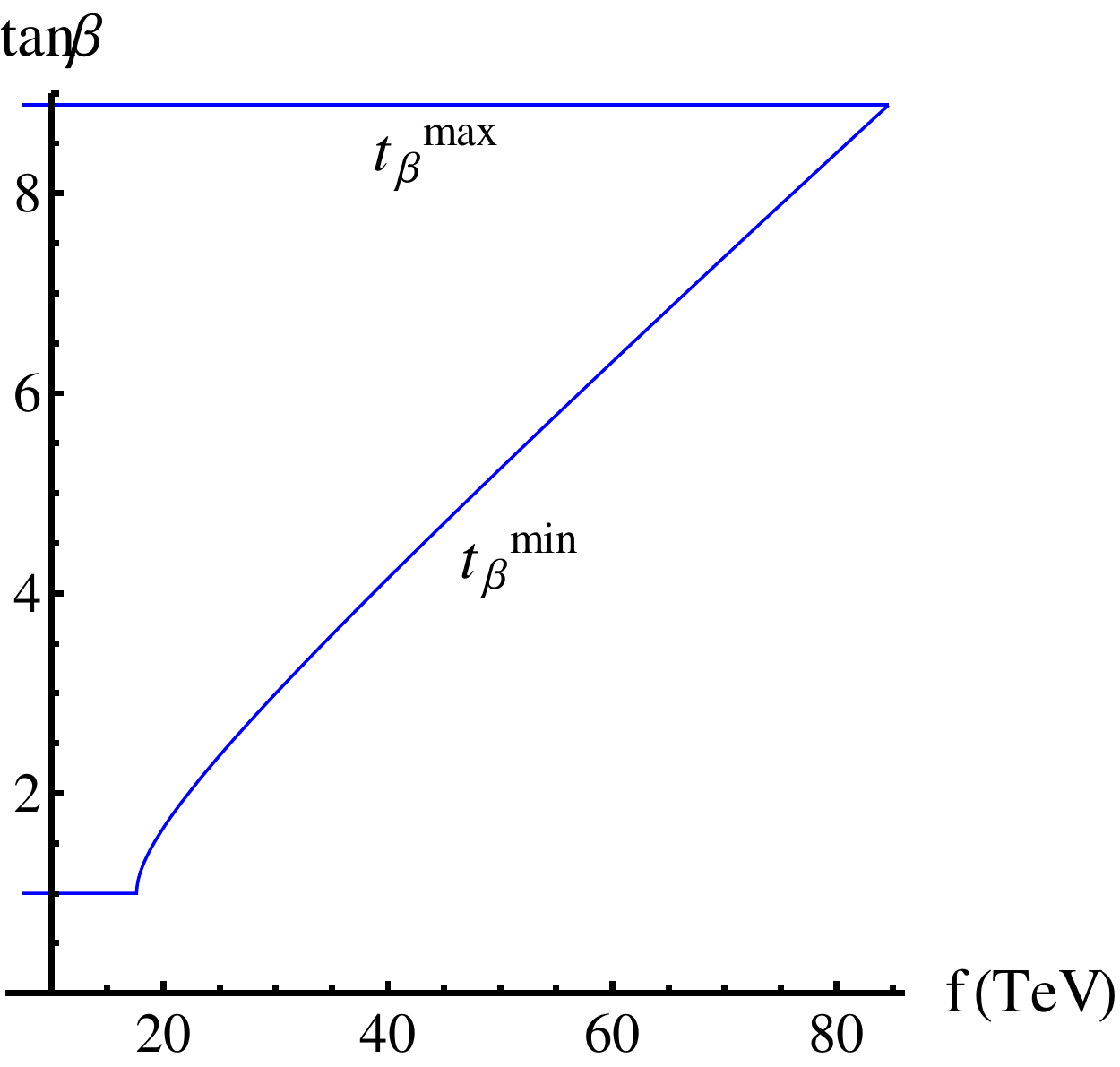}\quad\quad
\includegraphics[width=2.5in]{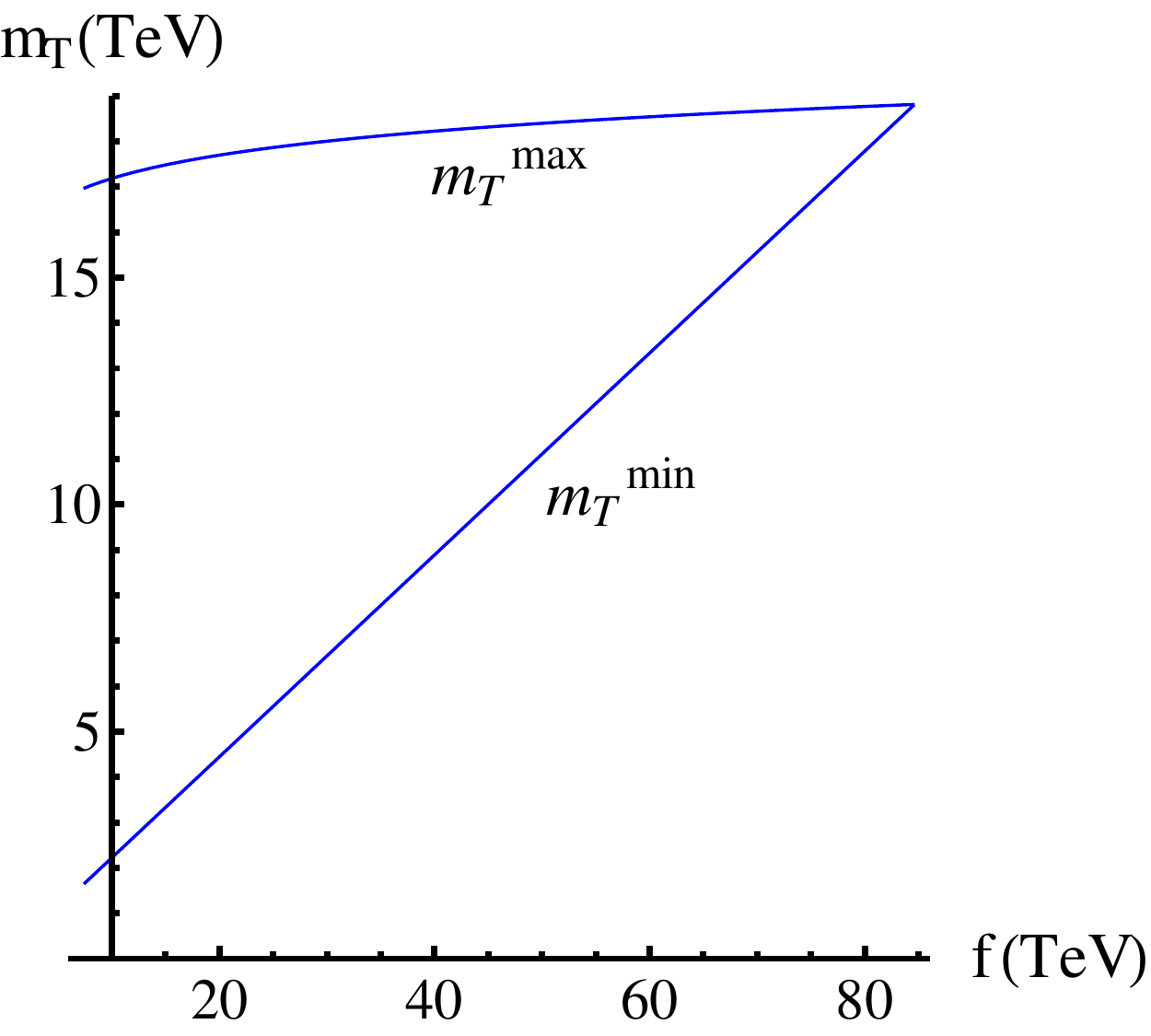}
\end{center}
\caption{\label{fig:minmax}Left: Maximum and minimum $t_\beta$ value as a function of $f$.
Right: Maximum and minimum $m_T$ value as a function of $f$.}
\end{figure*}
In figure~\ref{fig:minmax} we plot the maximum and minimum allowed $t_\beta$
and $m_T$ values as a function of $f$. Constraints from Eq.~\eqref{eq:poseta},
Eq.~\eqref{eq:MTmin}, Eq.~\eqref{eq:tbrange} and Eq.~\eqref{eq:mtu} are taken into account. From
the left panel of figure~\ref{fig:minmax} we see a constant $t_\beta^{\text{max}}$ value
($t_\beta^{\text{max}}\approx 9$) for all $f$. This follows from Eq.~\eqref{eq:tbrange} which requires $Z'$
mass should not exceed the unitarity constraint. Since $t_\beta$ is bounded
from above, Eq.~\eqref{eq:MTmin} then imposes a lower bound on top partner mass
for fixed $f$. This explains the $m_T^{\text{min}}$ curve shown in the right panel
of figure~\ref{fig:minmax}. The $m_T^{\text{max}}$ curve in the right panel of
figure~\ref{fig:minmax} is determined by Eq.~\eqref{eq:poseta}, reflecting the
fact that a too heavy top partner could lead to a negative pseudo-axion mass
squared. The upper bound on heavy top partner mass should be larger than its lower
bound, which leads to the increasing of $t_\beta^{\text{min}}$ for larger $f$
as shown in the left panel of figure~\ref{fig:minmax}. Therefore, with the increase
of $f$, the allowed range of variation for $t_\beta$ shrinks, and eventually
hit a point at $f\approx 85\TeV$, above which a perturbative treatment of the SLH
as an EFT might not be reliable. We remark here that it is important to retain
the field dependence in the field form factor $\Delta(\hat{h})$ in order to obtain reliable
estimates of $m_T^{\text{max}}$ and $m_\eta-m_T$ mass relation. Although the difference
in treating $\Delta(\hat{h})$ as field-independent and field-dependent is proportional
to a small quantity $A\approx 0.018$, the $M_T$ parameter enters as the argument
of a logarithmic function and is therefore very sensitive to such corrections. If
we treated $\Delta(\hat{h})$ as field-independent, we would have obtained
a maximum $m_T^{\text{max}}$ value of about half of the value obtained by retaining
field-dependence.

\begin{figure*}[ht]
\begin{center}
\includegraphics[width=2.8in]{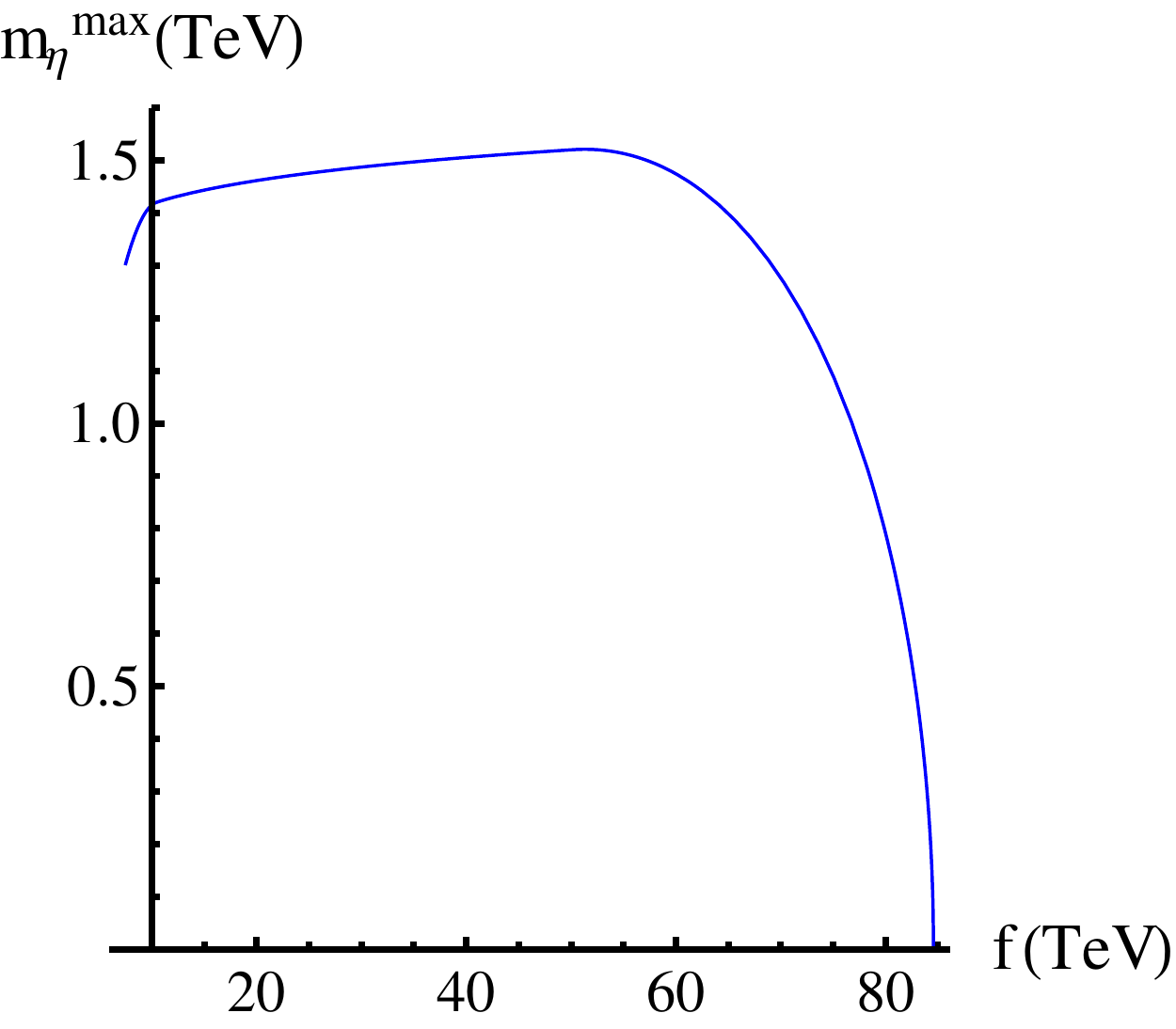}
\end{center}
\caption{\label{fig:metamax}Maximum $m_\eta$ value as a function of $f$.}
\end{figure*}
In figure~\ref{fig:metamax} we plot the maximum allowed $m_\eta$
value as a function of $f$. The minimum allowed $m_\eta$ value, before
any experimental or observational constraint is taken into account, is zero.
The maximum allowed $m_\eta$ for fixed $f$ reaches a maximum at about
$f\approx 51\TeV,m_\eta=1.5\TeV$.

\begin{figure*}[ht]
\begin{center}
\includegraphics[width=2.9in]{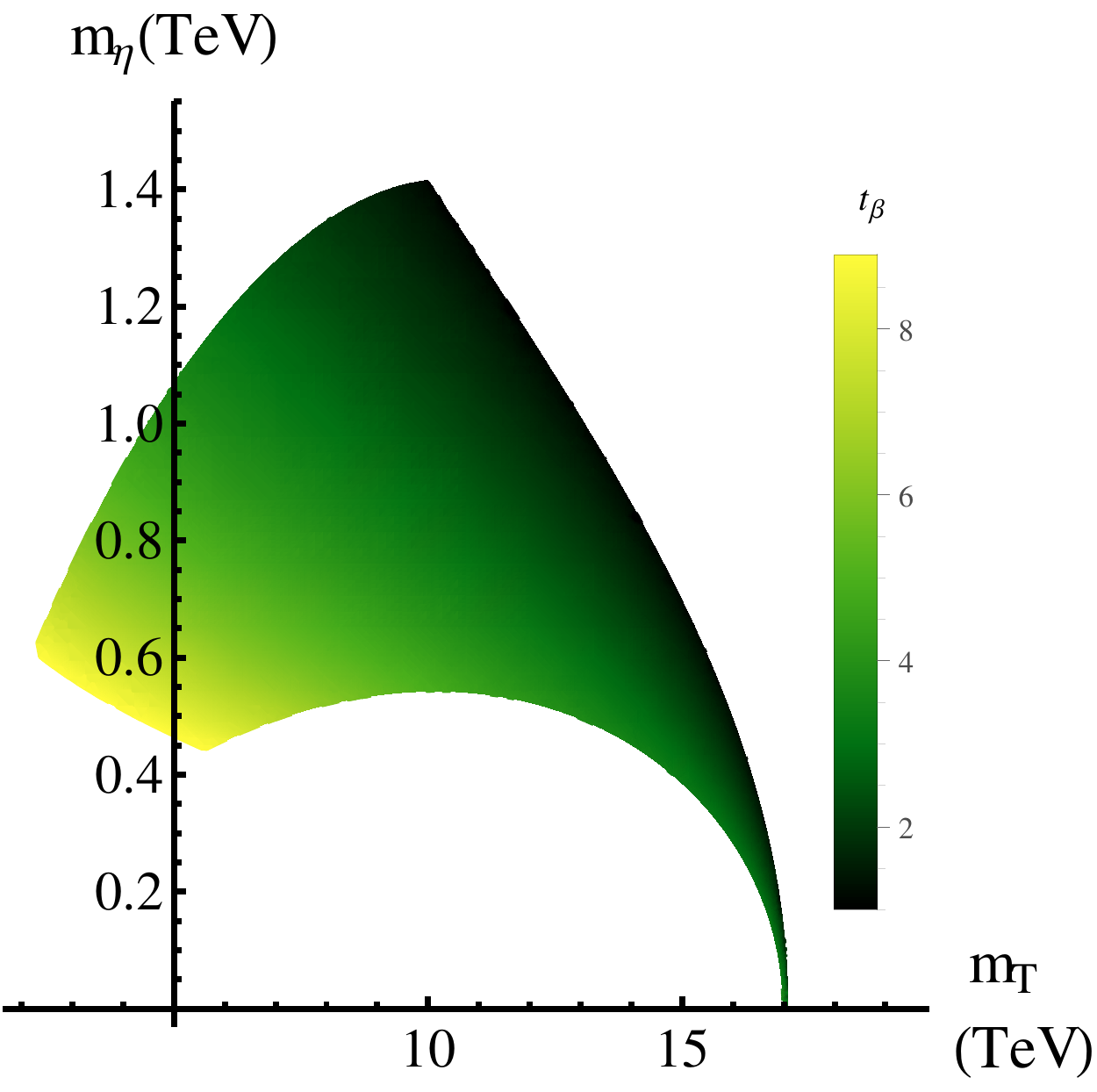}
\end{center}
\caption{\label{fig:tanb10}$t_\beta$ density plot in the allowed region
in the $m_\eta-m_T$ plane for $f=10\TeV$.}
\end{figure*}
\begin{figure*}[ht]
\includegraphics[width=1.9in]{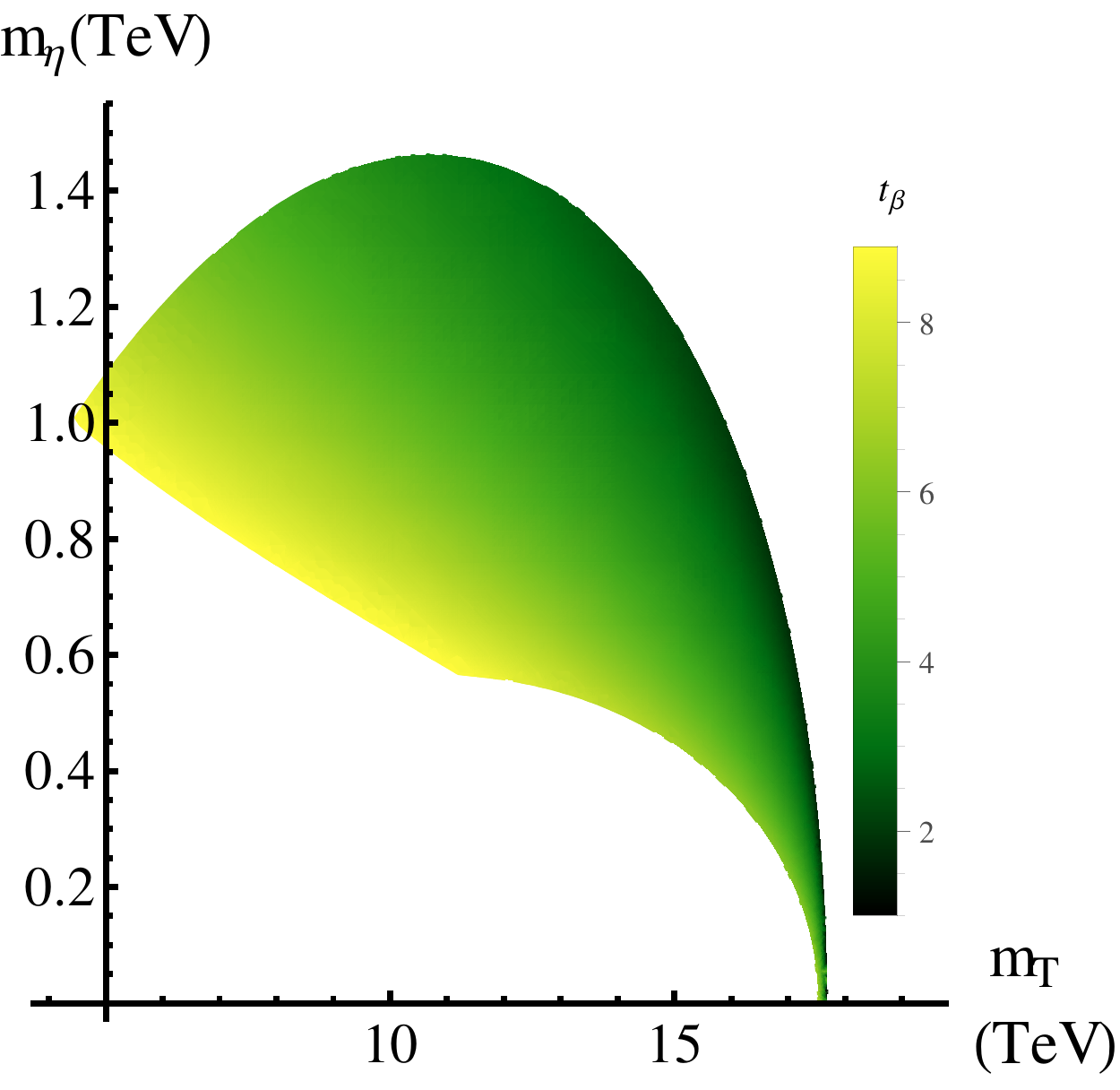}
\includegraphics[width=1.9in]{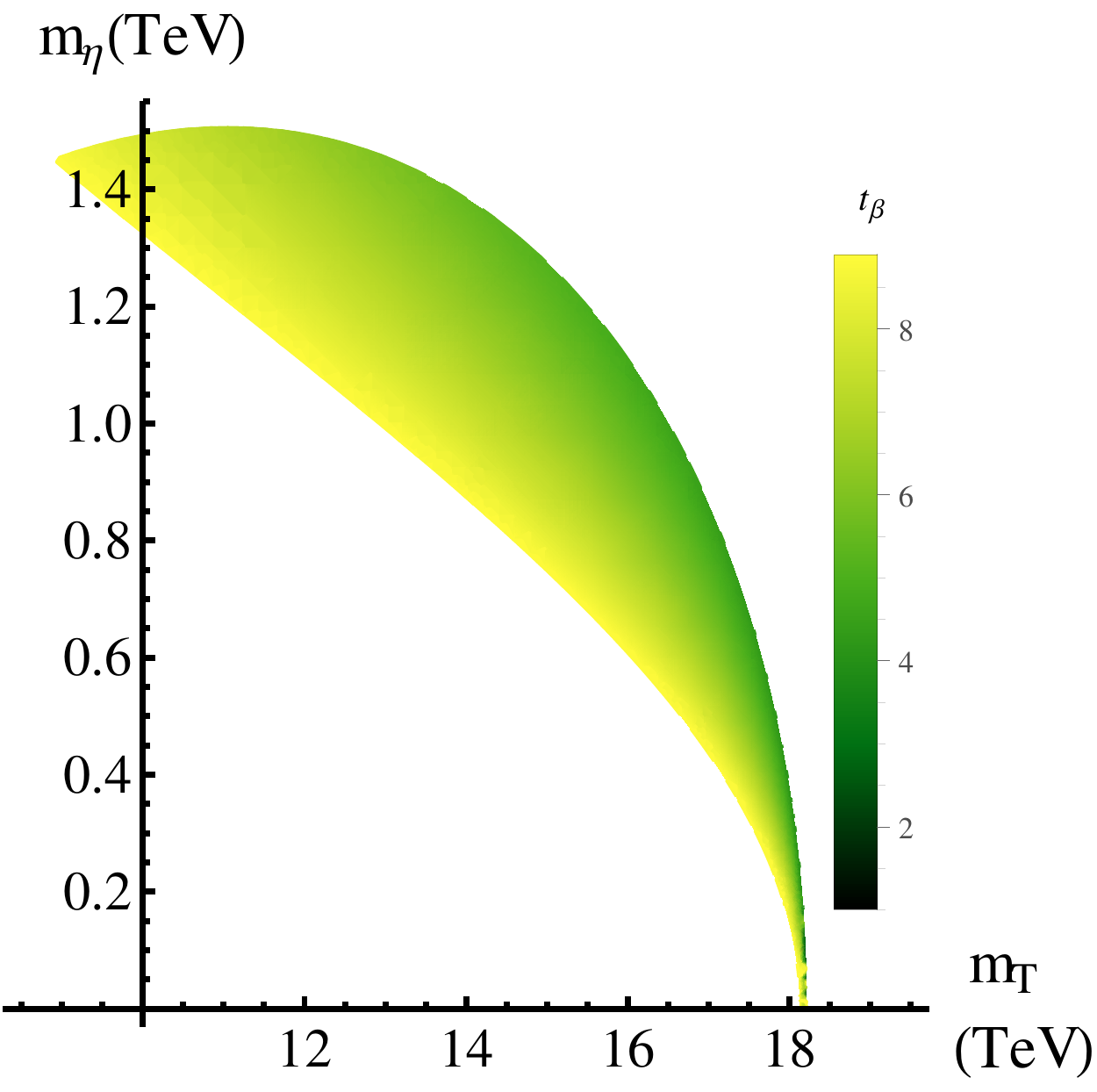}
\includegraphics[width=1.9in]{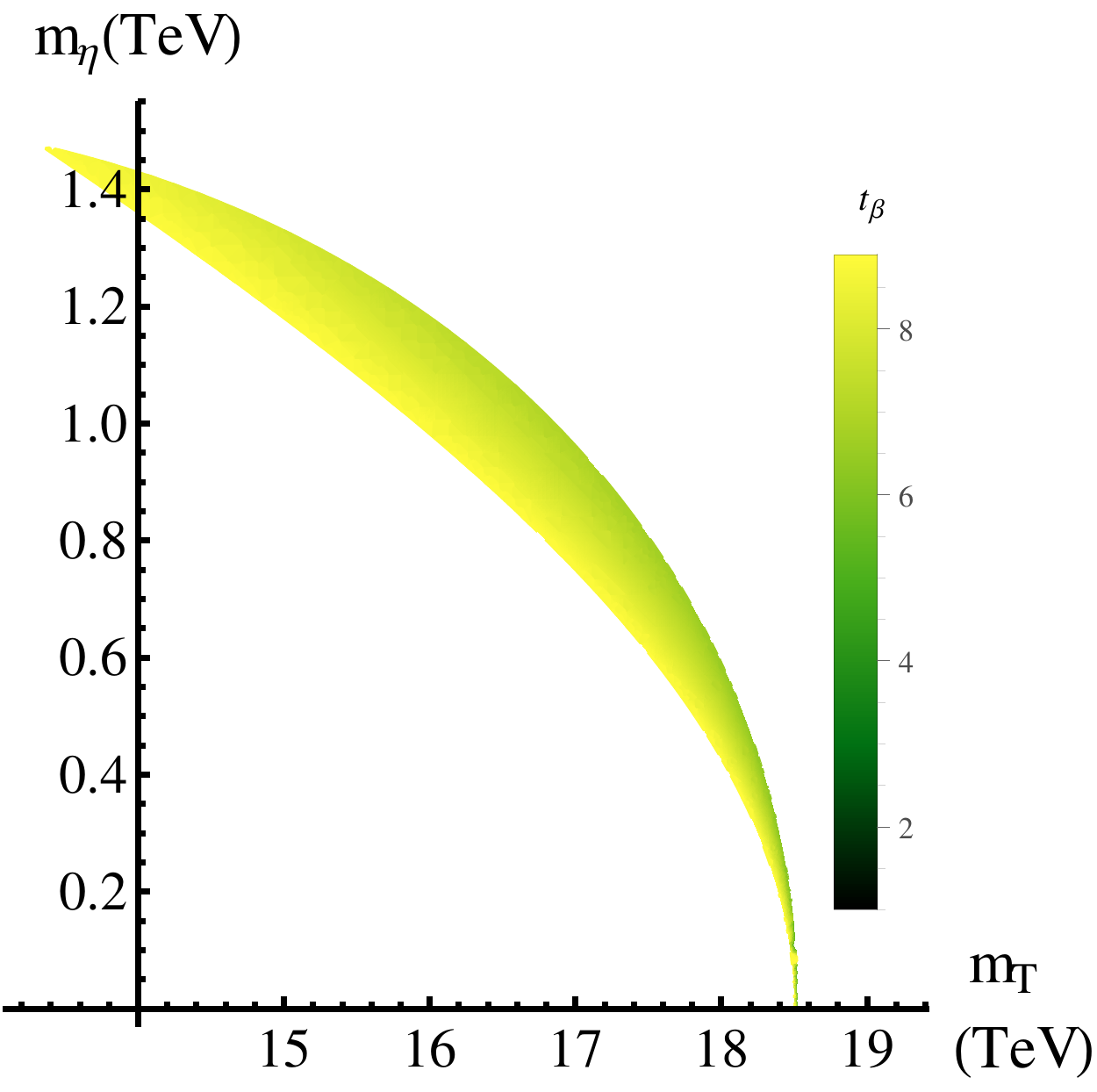}
\caption{\label{fig:tanb2070}$t_\beta$ density plot in the allowed region
in the $m_\eta-m_T$ plane for $f=20,40,60\TeV$.}
\end{figure*}
Figure~\ref{fig:tanb10} exhibits the $t_\beta$ value in the allowed region
in the $m_\eta-m_T$ plane for $f=10\TeV$. One easy way to understand this
figure is to follow the contour of constant $t_\beta$, which can be identified
by the color code. If we start from a high value of $t_\beta$
at some point with lighter color, then we can move along two opposite directions by
keeping $t_\beta$ fixed. We may move along the direction which increases
$m_T$ and eventually reach a point that saturates the unitarity bound
Eq.~\eqref{eq:mtu}, which is responsible for the dent in the lower part
of the figure. On the other hand, we may move along the direction which
decreases $m_T$ and eventually reach a point that saturates the bound in
Eq.~\eqref{eq:MTmin}, which determines the northwest boundary of the allowed
region. For $f=10\TeV$, the northeast and southwest boundary of the allowed
region are cut off by $t_\beta=1$ and $t_\beta=t_\beta^{\text{max}}\approx 9$,
respectively.

In figure~\ref{fig:tanb2070} we present the
$t_\beta$ density plot in the allowed region
in the $m_\eta-m_T$ plane for $f=20,40,60\TeV$. Because of
the increased $t_\beta^{\text{min}}$ value, the boundary
of the upper half allowed region will always be determined
by Eq.~\eqref{eq:MTmin} and it will not be cut off by $t_\beta=1$.
This explains the absence of a sharp turning point in the
upper boundary of the allowed region in figure~\ref{fig:tanb2070}.
Moreover, the effect of unitarity constraint on $m_T$
becomes milder, and vanishes for $f=40,60\TeV$ cases,
explaining the shrinking and disappearance of the dent
in the lower part.

\section{Naturalness in the Simplest Little Higgs}
\label{sec:nislh}

\subsection{Anatomy of Naturalness in Continuum Effective Field Theory}

Here we intend to make some clarification about the notion of naturalness
in accord with the spirit of CEFT, in order to lay the foundation
for our treatment of the degree of fine-tuning in the SLH.

Although the issue of Higgs mass naturalness~\cite{Wilson:1970ag,Gildener:1976ai,
Susskind:1978ms,tHooft:1979rat,Veltman:1980mj} has been the major
driving force of the development of TeV scale new physics model building for decades, unfortunately it
is often discussed in a manner which puzzles people and generates
confusion. One usual way to introduce this issue is by showing that
the one-loop correction to the Higgs boson mass in the SM is quadratically
divergent, and when we take the cutoff to be some high scale, say
the Planck scale or grand unification scale, then we need a huge
cancellation between the tree-level mass and one-loop correction
to obtain the observed light Higgs boson mass, implying a huge amount
of fine-tuning. Thus we need to introduce some new physics
to cancel the quadratic divergence in the one-loop correction to
the Higgs boson mass.

It does not require much reflection for one to smell something unusual from
the above usual argument. When we do QFT calculations, the regularization
cutoff is not an observable quantity, and should be removed after
renormalization. Then why should we worry about the fine-tuning originating
from this unobservable, unphysical quantity? This question was also raised
by R. Barbieri in ref.~\cite{Barbieri:2013vca}. To answer this question,
in ref.~\cite{Barbieri:2013vca} Barbieri tried to argue with only physical,
renormalized quantities. His argument is crucial and we would like to repeat
its main point here. Let us consider adding a real singlet scalar field
with a mass of around $M_H=10^{10}\GeV$ to the SM and coupling this scalar to
the SM Higgs doublet with strength $\lambda_H=1$. We may now examine the running
mass squared of the SM Higgs doublet $m_r^2$ and see how it evolves with scale. The result
is schematically shown in Figure 1. of ref.~\cite{Barbieri:2013vca}. Roughly speaking,
below $M_H=10^{10}\GeV$, $m_r^2$ remains around the electroweak scale.
At $M_H=10^{10}\GeV$, there is a turning point of $m_r^2$ evolution above which
$m_r^2$ fastly grow to order $(\lambda_H M_H^2)/(16\pi^2)$. The peculiar thing
about this running behavior is that $m_r^2$ at the electroweak scale is highly
sensitive to $m_r^2$ at a high scale ($>M_H$). In other words, the value of
$m_r^2$ at the high scale needs to be adjusted very delicately to obtain a low
value of $m_r^2$ at a scale below $M_H$. The smallness of our electroweak scale compared
to a high new physics scale $M_H$ is owing to a delicate adjustment of parameters
at the high scale. Similar things would happen if there are new fermions
or vector bosons at high scale which couple to the SM Higgs.

There are a number of issues which entail further clarification. Firstly, it is
clear from the above argument that, to formulate the Higgs mass naturalness
problem, new physics need to be introduced which couples to the SM Higgs. Pure
SM with its currently measured parameters does not suffer from the Higgs mass
naturalness problem~\cite{Wulzer:2015fya}. This is more clearly seen if we write down the renormalization
group equation (RGE) for the SM Higgs mass squared parameter in a mass-independent
renormalization scheme like $\overline{\text{MS}}$\footnote{For example, see ref.~\cite{Buttazzo:2013uya}.}.
Because it is the only dimensionful parameter in the SM
Lagrangian, it can only participate in a logarithmic running with the $\beta$
function proportional to itself. With the currently measured SM particle masses
this running behavior obviously does not exhibit a serious fine-tuning~\cite{Bardeen:1995kv}.
This certainly does not mean that the entire industry of searching for BSM
theories which alleviate the Higgs mass naturalness problem is useless. The reason
is that there certainly exist issues (e.g. the existence of dark matter, neutrino mass,
baryon asymmetry of the universe, strong CP problem, gravity, etc.) which are not
possible to explain (or be explained satisfactorily) in the pure SM. Extension
of the SM is not only a logical possibility, but also called for by the understanding
of the aforementioned issues and the understanding of the SM itself. However,
naive extensions as discussed in the previous paragraph will normally lead
to a fine-tuning problem, exhibited in the extreme sensitivity of IR parameters
to UV parameters in the theory. This is the real motivation to search for
mechanisms that protect the Higgs mass from being too sensitive to high
scale physics.

It is tempting to regard Barbieri's argument as only a more rigorous and refined formulation
of the Higgs mass naturalness problem compared to the conventional formulation
of investing the regularization cutoff with physical meaning. However, there is
indeed a crucial difference\footnote{This point is also emphasized by
ref.~\cite{deGouvea:2014xba}}. Taking the SM as an example, in the conventional
formulation, the radiative sensitivity of Higgs mass comes from the interaction
of SM particles interacting with the Higgs, with the largest contribution supposedly
from the top quark. However, by Barbieri's formulation, the radiative sensitivity
of Higgs mass should mainly come from the new physics which interacts with
the SM Higgs. Of course new physics need not interact with the Higgs boson with
the same strength as SM particles do. In the case that new physics only interacts
with the SM Higgs very weakly, the degree of fine-tuning in Barbieri's formulation
can be small and does not agree with the large fine-tuning expected from the
conventional formulation.

We note that one way to interpret the conventional formulation
seriously is to consider a Wilsonian formulation of QFT,
in which there is indeed an intrinsic cutoff defined in the functional integral.
In this framework the interpretation of fine-tuning crucially depends on
whether the cutoff is only a mathematical construct or it does have some
physical significance just as the atomic spacing in condensed matter systems.
Even if the cutoff does have physical significance, it is probably difficult (e.g. due
to the need to maintain certain symmetries) to draw a complete analogy between
the QFT describing the real particle phenomena and some condensed matter system,
when it comes to the way they connect to the underlying theory. Only in the case
that the cutoff indeed has physical significance can we make sense of the
conventional fine-tuning formulation in this framework. However, in low energy
experiments we are not able to determine whether the cutoff is physical. If the
cutoff is considered unphysical, then care must be taken when one tries to
extract fine-tuning information from a Wilsonian analysis. This is because in
a Wilsonian analysis the expression of the sensitivity of IR parameters to UV
parameters often invokes cutoff. If the cutoff is unphysical then further
efforts are required to disentangle physical fine-tuning\footnote{Here ``physical
fine-tuning'' means fine-tuning that is completely associated with physical parameters.} from the artificial
cutoff dependence (one such attempt is made in ref.~\cite{Aoki:2012xs}). Of course
when the smoke clears we would expect the physical fine-tuning is similarly
encoded in the logarithmic RGE behavior as dictated by its CEFT counterpart. The Wilsonian
viewpoint sheds light on the crucial difference between two fine-tuning formulations
pointed out in the previous paragraph, in that the difference can be traced back
to whether the cutoff has physical significance as opposed to only acting as
a mathematical regulator for the functional integral. Depending on the answer to
this question, different approaches to fine-tuning definition can be considered.
For example, ref.~\cite{Farina:2013mla} proposed the notion of ``finite naturalness''
in which the ``uncomputable'' power divergences are ignored. In the following
we opt to consider the cutoff as unphysical and only discuss the physical fine-tuning
in the theory.

In literature sometimes the Higgs mass naturalness problem is considered to be
equivalent to the question of why the Higgs mass (or equivalently, the electroweak scale)
is much smaller than some new physics scale, say Planck scale. There are two layers
of meaning in this question. The first layer can be called ``ensemble naturalness''.
It means that when we pose the naturalness question, we are implicitly referring to
an ensemble in which parameters of the theory are varied according to some
distribution~\cite{Hall:2007ja}. Then the naturalness question is about why some
parameters of the theory are very close to some special values. Sometimes physicists
consider $\ord(1)$ parameters as fully natural while extremely small parameters
(or large hierarchy of parameters) would need an explanation. In this sense, not only
the smallness of the ratio of Higgs mass to Planck scale needs an explanation, but also
the smallness of the electron Yukawa compared to top Yukawa would need an explanation.
Such an explanation seems to be available only when we know the UV theory which contains
the mechanism needed to explain the observed SM parameters. The second layer of meaning
is ``technical naturalness''. In technical naturalness, the absolute smallness
of some parameter is not regarded as a problem \textit{a priori}. However, if the
smallness of some IR parameters is not stable against tiny variation of UV parameters (i.e.
parameters defined at a high scale) of the theory, then the problem arises as to
why we are in such an unstable parameter configuration. 't Hooft suggested that if the
vanishing of a small parameter would enhance the symmetry of the theory, then it could
be considered as technically natural~\cite{tHooft:1979rat}. Therefore, the smallness of electron Yukawa
is not considered as a naturalness problem in this sense, since the vanishing of
the electron Yukawa would restore the associated chiral symmetry. More generally,
if the smallness of some parameters are protected by the renormalization structure
of the theory, then the smallness can be seen as technically natural. Compared to
ensemble naturalness, the discussion of technical naturalness is more convenient,
because technical naturalness can be examined even if we know very little about the
UV completion of the theory and only have knowledge about the low energy effective
theory. In the present paper, the Higgs mass naturalness problem that we aim to
discuss clearly belongs to technical naturalness.

Let us return to Barbieri's example of SM plus a large mass scalar with the aim of
understanding its technical (un)naturalness. As noted before, if there is only
the SM, then there would be no Higgs mass naturalness problem. The logarithmic
running of the scalar mass squared parameter in the pure SM does not show a sensitive character
for a long range of renormalization group (RG) flow. The real cause of the sensitivity
of the IR parameter to the UV parameter here is a special turning point($M_H\sim10^{10}\GeV$)
in the RG flow. We say the turning point is special in that the speed of the running
of $m_r^2$ exhibits a huge difference below and above the turning point, which is the
direct cause of the extreme sensitivity. There are also turning points if we investigate
the running of gauge couplings. For example, the running behavior of $\alpha_S$ changes
when a quark threshold is crossed. Nevertheless, the change is mild since it is not tied
to the power of scale and therefore cannot lead to a serious fine-tuning. The situation
is obviously different for dimensionful parameters which could mix with other high scales
in the theory.

In Barbieri's example, the appearance of the turning point in the RG flow can be traced
back to the fact that below the threshold ($M_H\sim10^{10}\GeV$), the large mass scalar
has been integrated out and only the SM fields contribute to the RG, consistent with the
expectation from the Appelquist-Carazzone decoupling~\cite{Appelquist:1974tg}. Moreover, it should
be noted that implicit in this example is the use of some mass-independent renormalization
scheme (say $\overline{\text{MS}}$) both below and above the threshold to express the RGE.
These theoretical treatments are reflection of the spirit of the CEFT which suggests
describing the physics associated with a certain energy scale with descriptions appropriate
to that energy scale. The continuum RGE certainly embodies this spirit and gives rise to
important physical predictions, such as asymptotic freedom for QCD. However, the mass-independent
RGE alone certainly is not the whole story, since a naive crossing of a mass threshold
with no modification of the mass-independent RGE amounts to adding up the wrong infinite
set of logarithms and thus fails to describe the physics (i.e. Appelquist-Carazzone decoupling) faithfully~\cite{Georgi:1994qn}.
As emphasized in ref.~\cite{Georgi:1994qn}, the best solution is to use CEFT, in which
the Appelquist-Carazzone decoupling is put in by hand,
which is the heart of CEFT. The CEFT approach therefore
provides a simple framework for the demonstration of physical fine-tuning present
in the theory.

We are now prepared to analyze the fine-tuning problem in a general manner within
the CEFT framework. A global physical picture suggested by CEFT is that between thresholds,
the renormalization structure of a QFT is encoded in its RGE in a mass-independent
renormalization scheme, while when going below a mass threshold, heavy degrees of freedom
should be integrated out and the switch to a low energy effective QFT should be made
in order to facilitate an easy grasp of the main feature of the theory. Accordingly,
two sources of physical fine-tuning in this global picture can be identified. Firstly,
there can be fine-tuning associated with the RG running between thresholds (referred to as
``RG tuning''). This happens
if the value of a running parameter is very small at a particular renormalization scale
for observation while its running is very fast (i.e. $\beta$ function is large). In such
a case when we consider a finite range of RG flow then it is obvious that there can be
a high sensitivity of the IR parameter to the UV parameter. A simple example is scalar
QED, in which if a scalar quartic interaction is not written down, it can still be generated
via RG running, and therefore requiring the quartic to vanish at a particular point
would be a type of fine-tuning. We note that such kind of fine-tuning in the context of
Little Higgs models has been pointed out before~\cite{Grinstein:2009ex}. Secondly, there
can be fine-tuning associated with the transition from the EFT above the threshold to
the EFT below the threshold (referred to as ``threshold tuning''). Generally speaking,
the degrees of freedom and the description of the theory change due to crossing the
threshold. This change of decription could induce additional sensitivity of IR
parameters to UV parameters. We will see at below how the threshold tuning in the SLH
can be identified. However, we note that in the Barbieri's example, the source of tuning
is RG tuning rather than threshold tuning. The reason is that $m_r^2$ can be seen as
continuous at the turning point and no threshold tuning is there. The real tuning is due to
a very small $m_r^2$ value at the threshold while the RG running above the threshold is
very fast and therefore belongs to RG tuning.

\subsection{Fine-Tuning in the SLH}

The SLH can be considered an EFT valid up to its unitarity cutoff $\Lambda_U\equiv\sqrt{8\pi}fc_\beta$.
Roughly speaking, there is a scale separation in the validity range of the SLH.
Two characteristic scales can be identified, one is associated with the scales of
heavy sector particles, such as $m_T, m_{Z'}$, the other is the electroweak scale
which can be represented by $v$ or $m_h$. The fine-tuning in the SLH, from the CEFT
point of view, is about how the electroweak scale parameters, such as $v$ or $m_h$,
are sensitive to high scale parameters in the model. Here the ``high scale'' is
naturally chosen to be the highest scale at which the SLH claims to be valid as
an EFT. In our analysis this is unsurprisingly chosen to be the unitarity cutoff
$\Lambda_U$. For simplicity we ignore small input parameter corrections in the
fine-tuning analysis.

The values of $v$ and $m_h$ can be calculated once the following set of parameters
are given at $\Lambda_U$: $f,t_\beta,M_T,\lambda_R,\mu^2$. In our analysis we neglect
field strength renormalization effects and the only running parameter in this set
is $\lambda_R$. The procedure to calculate $v$ and $m_h$ is then straightforward:
we first follow the RG flow to a scale $M_L$ which satisfies $\bar{\lambda}=0$
at $\mu_R=M_L$ ($\bar{\lambda}$ is defined by Eq.~\eqref{eq:blambda}). At $\mu_R=M_L$
we have the parameter $\lambda$ introduced in Eq.~\eqref{eq:vrc} just given
by $\lambda_R$. Next we may use the analysis in the previous section to obtain
$v$ and $m_h$, which can be viewed as transition to a low energy EFT for the
electroweak scale. The RG tuning and threshold tuning can be obtained respectively
from the above two steps.

Let us first investigate the threshold tuning. This requires us to express $v$ and
$m_h$ in terms of $\lambda,\mu^2$ and $f,t_\beta,M_T$. From the stationary point
condition Eq.~\eqref{eq:spc} we could obtain the following expression of $v^2$
by expanding the sines to $\hat{h}_0^3$
\begin{align}
v^2=\frac{4\left(\lambda f^2-\frac{\mu^2}{s_\beta c_\beta}\right)}
{\frac{1}{3f^2 s_\beta^2 c_\beta^2}\left(4\lambda f^2-\frac{\mu^2}{s_\beta c_\beta}\right)
+4\Delta_0-2A}
\label{eq:v24ft}
\end{align}
in which $\Delta_0$ and $A$ are given by Eq.~\eqref{eq:delta0} and
Eq.~\eqref{eq:A}, respectively. Then we may expand the cosines in
Eq.~\eqref{eq:mh2} to $\hat{h}_0^2$ and plug in Eq.~\eqref{eq:v24ft}
to obtain
\begin{align}
m_h^2=2\left(\lambda f^2-\frac{\mu^2}{s_\beta c_\beta}\right)-2Av^2
\label{eq:mh24ft}
\end{align}
in which $v^2$ is given by Eq.~\eqref{eq:v24ft}. Let us define
threshold tuning $\Delta_{\TH}^\lambda,\Delta_{\TH}^{\mu^2}$ for
$\lambda$ and $\mu^2$ as follows (with inspiration from ref.~\cite{Barbieri:1987fn})
\begin{align}
\Delta_{\TH}^\lambda\equiv\left|\frac{\lambda}{m_h^2}\frac{\partial m_h^2}{\partial \lambda}\right| \\
\Delta_{\TH}^{\mu^2}\equiv\left|\frac{\mu^2}{m_h^2}\frac{\partial m_h^2}{\partial \mu^2}\right|
\end{align}
These definitions obviously reflect how the relative variation
of $m_h^2$ is sensitive to the relative variation of $\lambda$ and $\mu^2$.
From Eq.~\eqref{eq:mh24ft} the threshold tuning values are calculated to be
\begin{align}
\Delta_{\TH}^\lambda & =1+\frac{2m_\eta^2}{m_h^2} \\
\Delta_{\TH}^{\mu^2} & =\frac{2m_\eta^2}{m_h^2}
\end{align}
In obtaining the above threshold tuning values we have neglected
terms proportional to $A$ or relatively suppressed by $\frac{v^2}{f^2}$.
Also, we checked that if we use $v^2$ instead of $m_h^2$ to quantify
the threshold tuning, the results do not change. The above results
suggest that the threshold tuning is determined by $m_\eta$: with
larger $m_\eta$ we get larger tuning. This is easy to understand
since from Eq.~\eqref{eq:mh24ft} we see that $m_h^2$ can be approximately
viewed as the result of cancellation between $2\lambda f^2$ and
$\frac{2\mu^2}{s_\beta c_\beta}\sim 2m_\eta^2$.

To calculate the RG tuning we need the $\beta$ function of $\lambda_R$,
denoted as $\beta_\lambda$. In this work we neglect the contribution
to $\beta_\lambda$ from field strength renormalization and scalar loop.
From Eq.~\eqref{eq:bl} the following expression for $\beta_\lambda$
can be derived
\begin{align}
\beta_\lambda=-\frac{3\lambda_t^2}{4\pi^2}\frac{M_T^2}{f^2}
+\frac{3g^4}{32\pi^2}\frac{5+t_W^2}{3-t_W^2}
\end{align}
We use $\lambda_U$ to denote the value of $\lambda_R$ defined
at the unitarity cutoff $\Lambda_U$. Then the relation between $\lambda$ and
$\lambda_U$ can be expressed as
\begin{align}
\lambda=\lambda_U-\beta_\lambda\ln\frac{\Lambda_U}{M_L}
\end{align}
The RG tuning of $\lambda$ is then defined through
\begin{align}
\Delta_\RG^\lambda\equiv\left|\frac{\lambda_U}{\lambda}
\frac{\partial\lambda}{\partial\lambda_U}\right|
=\left|1+\frac{1}{\lambda}\beta_\lambda\ln\frac{\Lambda_U}{M_L}\right|
\label{eq:drgldf}
\end{align}
The special scale $M_L$ is defined by making $\bar{\lambda}$ vanish. From
this requirement we find the following useful relation for
analysis of the RG tuning
\begin{align}
\left[\lambda_t^2 M_T^2-\frac{5+t_W^2}{8(3-t_W^2)}g^4 f^2\right]
\ln\frac{M_T}{M_L}=\frac{1}{2}B
\end{align}
where $B$ is defined as
\begin{align}
B\equiv\lambda_t^2 M_T^2+\frac{1}{4}g^2 M_X^2\left(
\ln\frac{M_X^2}{M_T^2}-\frac{1}{3}\right)+\frac{1}{8}g^2 (1+t_W^2)
M_{Z'}^2\left(\ln\frac{M_{Z'}^2}{M_T^2}-\frac{1}{3}\right)
\end{align}
$\Delta_\RG^\lambda$ is then calculated to be ($\Lambda_U=\sqrt{8\pi}fc_\beta$)
\begin{align}
\Delta_\RG^\lambda=\left|1-\frac{3}{2\pi^2}\frac{\lambda_t^2 M_T^2
-\frac{g^4 f^2}{8}\frac{5+t_W^2}{3-t_W^2}}{m_h^2+2m_\eta^2}\ln
\frac{\sqrt{8\pi}fc_\beta}{M_T}-\frac{3}{4\pi^2}\frac{B}{m_h^2+2m_\eta^2}\right|
\end{align}
To obtain this expression we have made use of Eq.~\eqref{eq:mh24ft}
to find the $\lambda$ expressed by $m_h$ and $m_\eta$ and again
neglected corrections suppressed by $\frac{v^2}{f^2}$ or $A$.
\begin{figure*}[ht]
\begin{center}
\includegraphics[width=3.0in]{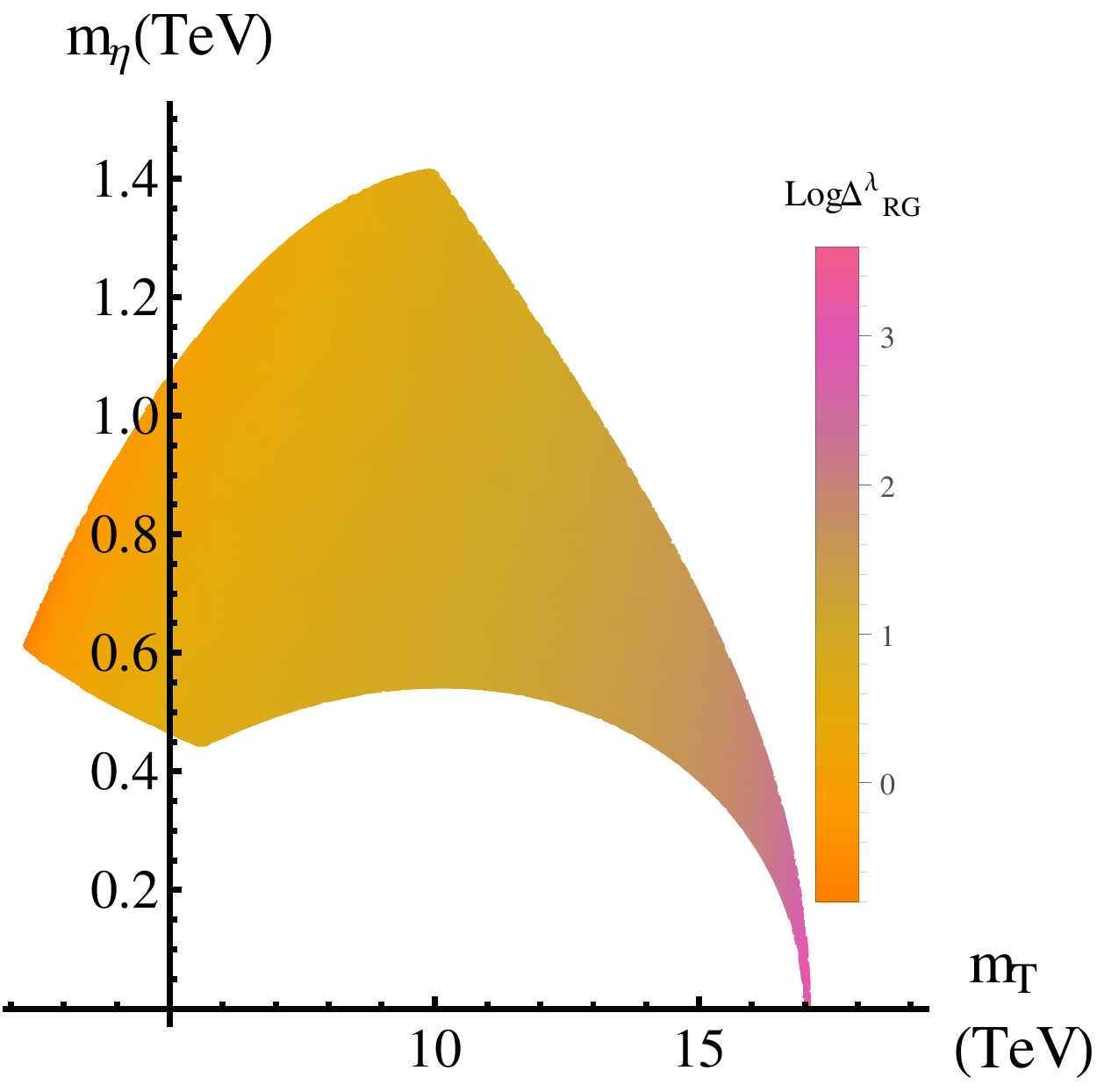}
\end{center}
\caption{\label{fig:drge} Density plot of $\text{Log}\Delta_\RG^\lambda$ in the
$m_\eta-m_T$ plane for $f=10\TeV$. Here $\text{Log}$ means $\log_{10}$.}
\end{figure*}
Figure~\ref{fig:drge} presents the density plot of $\text{Log}\Delta_\RG^\lambda$ in the
$m_\eta-m_T$ plane for $f=10\TeV$. We note that in the left region where
$m_T$ is light, $\Delta_\RG^\lambda$ can drop below $1$, but not reach $0$.
We have checked that for $f=10\TeV$, $\lambda_U$ is always negative.
Figure~\ref{fig:drge} can be regarded alternatively as an indication of how
negative $\lambda_U$ is. For large $\Delta_\RG^\lambda$ it is therefore
natural to question the vacuum stability of the corresponding parameter
point. Because there still exists parameter region of relatively small
$\Delta_\RG^\lambda$ we do not expect in any case that vacuum stability consideration should
exclude all the allowed parameter space in figure~\ref{fig:drge}, although
a detailed study is beyond the scope of the present paper.

The RG tuning of $\mu^2$ parameter denoted as $\Delta_{\RG}^{\mu^2}$ can also
be discussed, but this turns out to be trivial, i.e.
\begin{align}
\Delta_{\RG}^{\mu^2}=1
\label{eq:drgmu2}
\end{align}
since the $\mu_R^2$ does not run (when wave-function renormalization and
small contribution from light Yukawa are neglected).

The threshold tuning and RG tuning derived above can be combined to define
a total tuning of a given parameter (e.g. $\lambda$ or $\mu^2$). For example,
the total tuning of $\lambda$ is defined as
\begin{align}
\Delta_{\TOT}^\lambda\equiv\Delta_{\TH}^\lambda\times\Delta_{\RG}^\lambda
\end{align}
The use of multiplication in the above definition is easy to understand: the
total tuning defined in this way just reflects how the relative change of $m_h^2$
is sensitive to the relative change of $\lambda_U$. For $\mu^2$, we have
\begin{align}
\Delta_{\TOT}^{\mu^2}\equiv\Delta_{\TH}^{\mu^2}\times\Delta_{\RG}^{\mu^2}
=\Delta_{\TH}^{\mu^2}
\end{align}
where the second step is due to Eq.~\eqref{eq:drgmu2}. Finally, to quantify the
overall degree of fine-tuning in the SLH, we define
\begin{align}
\Delta_{\TOT}=\max\{\Delta_{\TOT}^{\mu^2},\Delta_{\TOT}^\lambda\}
\end{align}
For simplicity we do not attempt a more sophisticated statistical combination.

Let us take a closer look at $\Delta_{\TOT}^\lambda$, which is easily
calculated to be
\begin{align}
\Delta_{\TOT}^\lambda=\left|1+\frac{2m_\eta^2}{m_h^2}-\frac{3}{2\pi^2}\frac{\lambda_t^2 M_T^2
-\frac{g^4 f^2}{8}\frac{5+t_W^2}{3-t_W^2}}{m_h^2}\ln
\frac{\sqrt{8\pi}fc_\beta}{M_T}-\frac{3}{4\pi^2}\frac{B}{m_h^2}\right|
\label{eq:dtotl}
\end{align}
It is worth noticing that part of the above equation (the term containing
$\ln\frac{\sqrt{8\pi}fc_\beta}{M_T}$) is very similar to the fine-tuning definition
employed in ref.~\cite{Reuter:2012sd}, which we copy here for convenience
\footnote{A similar definition is also employed by ref.~\cite{Yang:2014mba} in the context
of the Littlest Higgs model with T-parity.}
(see Eq.(4.2) and Eq.(4.3) in ref.~\cite{Reuter:2012sd})
\begin{align}
\Delta=\frac{|\delta\mu^2|}{\mu_{\text{obs}}^2},\mu_{\text{obs}}^2=\frac{m_h^2}{2},
\delta\mu^2=-\frac{3\lambda_t^2 m_T^2}{8\pi^2}\log\frac{\Lambda^2}{m_T^2}
\label{eq:oft}
\end{align}
In the above equation the notations are in accord with ref.~\cite{Reuter:2012sd}.
Furthermore, ref.~\cite{Reuter:2012sd} uses $\Lambda=4\pi f$ to cut off the
divergent one-loop integral. We see that if in our expression of
$\Delta_{\TOT}^\lambda$, we only retain the term containing
$\ln\frac{\sqrt{8\pi}fc_\beta}{M_T}$, neglect the
$\frac{g^4 f^2}{8}\frac{5+t_W^2}{3-t_W^2}$ part, and let
$\Lambda_U=\sqrt{8\pi}fc_\beta\rightarrow\Lambda=4\pi f$, then
we recover the fine-tuning definition of ref.~\cite{Reuter:2012sd}.
The differences between our treatment and the definition in ref.~\cite{Reuter:2012sd}
deserves some comments. The use of unitarity cutoff $\Lambda_U$ instead of
NDA cutoff $\Lambda$ is not essential, although this could have some
impact on the quantitative value of fine-tuning. The
$\frac{g^4 f^2}{8}\frac{5+t_W^2}{3-t_W^2}$ part reflects the interpretaional
difference between our treatment and that of ref.~\cite{Reuter:2012sd}.
In ref.~\cite{Reuter:2012sd}, the regularization cutoff is invested
with a physical meaning and the different terms in the one-loop $\delta\mu^2$
expression are then considered as independent sources of tuning. However,
in our treatment, this expression reflects the RG running
of $\lambda_R$ and through this running we infer how the IR parameters
are sensitive to the UV parameters. It is then clear that the
$\frac{g^4 f^2}{8}\frac{5+t_W^2}{3-t_W^2}$ part should be retained
because it also contributes to the $\beta$ function. Since this part
has a relative minus sign compared to the fermionic contribution,
it effectively reduces the fine-tuning in the model. In Eq.~\eqref{eq:dtotl}
we also have a $\frac{3}{4\pi^2}\frac{B}{m_h^2}$ term which is absent
in the definition of ref.~\cite{Reuter:2012sd}. This is again due to
the interpretational difference. In our treatment, the starting point of
the RG running is $M_L$, and only when the fermionic contribution
is much larger than the gauge contribution to the scalar effective potential
can we have $M_L\sim M_T$. Finally, in Eq.~\eqref{eq:dtotl}
we have the $1+\frac{2m_\eta^2}{m_h^2}$ part which is also absent in
ref.~\cite{Reuter:2012sd}. We can understand the role of this part in the
following manner. Suppose we turn off the $1+\frac{2m_\eta^2}{m_h^2}$ part
for the moment, then we could realize that the situation would
correspond to when defining the RG tuning $\Delta_{\RG}^\lambda$
we omit the ``1'' before the plus sign in Eq.~\eqref{eq:drgldf}.
In other words, the ratio of the amount of RG running to
the value of $\lambda_R$ defined at $M_L$ is taken to be the
measure of RG tuning. Intuitively this seems to be fine, however
this is different from the notion of the sensitivity of IR parameters
to UV parameters. Especially if we consider the case when $\beta_\lambda=0$,
using the definition with ``1'' omitted would lead to zero
RG tuning. Nevertheless if we think about the parameter sensitivity,
we would still realize that there is a $100\%$ sensitivity here,
i.e. when $\lambda_U$ changes a fraction then $\lambda$ would change
exactly the same fraction. Therefore, if we want to use
consistently the definition of various fine-tuning measures
as a reflection of sensitivity of IR parameters to UV parameters,
we should retain the ``1'' in the definition of RG tuning.

The above discussion in fact leads to a more clear understanding
of the NDCC assumption. In previous literature, this is equivalent
to calculating the scalar effective potential
by turning off the relevant tree-level contribution and imposing a
NDA cutoff to momentum integral. In our fine-tuning analysis we see that
for those parameter points that satisfy $\lambda_U=0$, the total tuning
of $\lambda$, denoted as $\Delta_{\TOT}^\lambda$ vanishes since
the RG tuning $\Delta_{\RG}^\lambda$ vanishes. Therefore, if we
interpret the NDCC assumption as corresponding to $\lambda_U=0$
in the CEFT approach, then parameter points that satisfy
this assumption will automatically have the property of making
$\Delta_{\TOT}^\lambda=0$. It is tempting to consider these
parameter points satisfying the NDCC assumption as particularly
good ones. However, the real situation is not that simple, due to
the following reasons. First, even if $\Delta_{\TOT}^\lambda=0$,
we still need to consider $\Delta_{\TOT}^{\mu^2}$, which cannot
be made arbitrarily small given $\Delta_{\TOT}^\lambda=0$. Second,
$\Delta_{\TOT}^\lambda=0$ is, honestly speaking, illusory. This is
because $\Delta_{\TOT}^\lambda=0$ is derived from a completely
IR point of view. If we think a little bit about UV completion,
then from the UV point of view, any tiny variation of UV fixed
that is possible to vary $\lambda_U$ slightly indicates an infinite
sensitivity and thus an infinite amount of fine-tuning. The problem
is that we choose a ``bad'' transition point to connect IR and UV.
If we use another scale at which $\lambda_U\neq 0$, we would be
able to obtain finite fine-tuning result.

It is also instructive at the moment to consider the alternative
interpretation~\cite{Casas:2005ev} that $\Lambda$ is chosen to be the renormalization
scale in conjunction with the NDCC assumption adopted
in previous literature. In our CEFT approach the renormalization scale
is denoted as $\mu_R$ (Eq.~\eqref{eq:bl}) and physics does not
depend on the choice of $\mu_R$. In fact we obtained the $\beta$ function
associated with $\lambda$ and the renormalized coupling $\lambda_R$
is required to have the correct $\mu_R$ dependence to cancel
the $\mu_R$ dependence in $\bar{\lambda}$. Then for fixed chosen
cutoff, the solution under the NDCC assumption obtained
by previous literature would correspond to a subset of the solution
obtained in our CEFT approach. The description of this
subset of solution, from the CEFT point of view, is a little bit subtle
since it \textit{seems} to violate closure under renormalization. However,
if we pick any one element of this subset, it does not violate RG invariance:
at the cutoff $\lambda_R$ vanishes and only receives loop corrections,
while at scale below cutoff $\lambda_R$ receives both tree and loop contribution.
The CEFT approach makes this point clear. However, we note that this alternative
interpretation can only hold when the replacement $\Lambda\rightarrow\mu_R$ indeed
yields the corresponding CEFT result. In those circumstances which terms involving
$\Lambda^2$ are retained, it is difficult to reconcile this interpretation with
a mass-independent renormalization scheme.

Therefore it is clarified that the commonly-seen NDCC assumption
actually corresponds to selecting a subset of parameter space obtained
via the CEFT approach, which satisfies $\lambda_U=0$.
We need to be careful about such selection, for the
following reasons. First, we need to check whether such selection is
consistent with the requirement of correct EWSB, given the measured electroweak
vacuum expectation value and Higgs mass. Second, there is no \textit{a priori}
reason to confine us to this selection. The nature may well allow some
amount of $\Delta_{\TOT}^\lambda$. Third, the real EFT cutoff is
unknown, and the physical predictions made based on some fixed $\Lambda$
value could be unreliable. This is the crucial place where CEFT reveals its power:
as long as we accept the SLH as a self-contained low energy EFT where
perturbation theory is valid, we are able to establish the mass relation
like Eq.~\eqref{eq:mr2} without reliance on the knowledge or assumption
about the cutoff.

\begin{figure*}[ht]
\begin{center}
\includegraphics[width=2.5in]{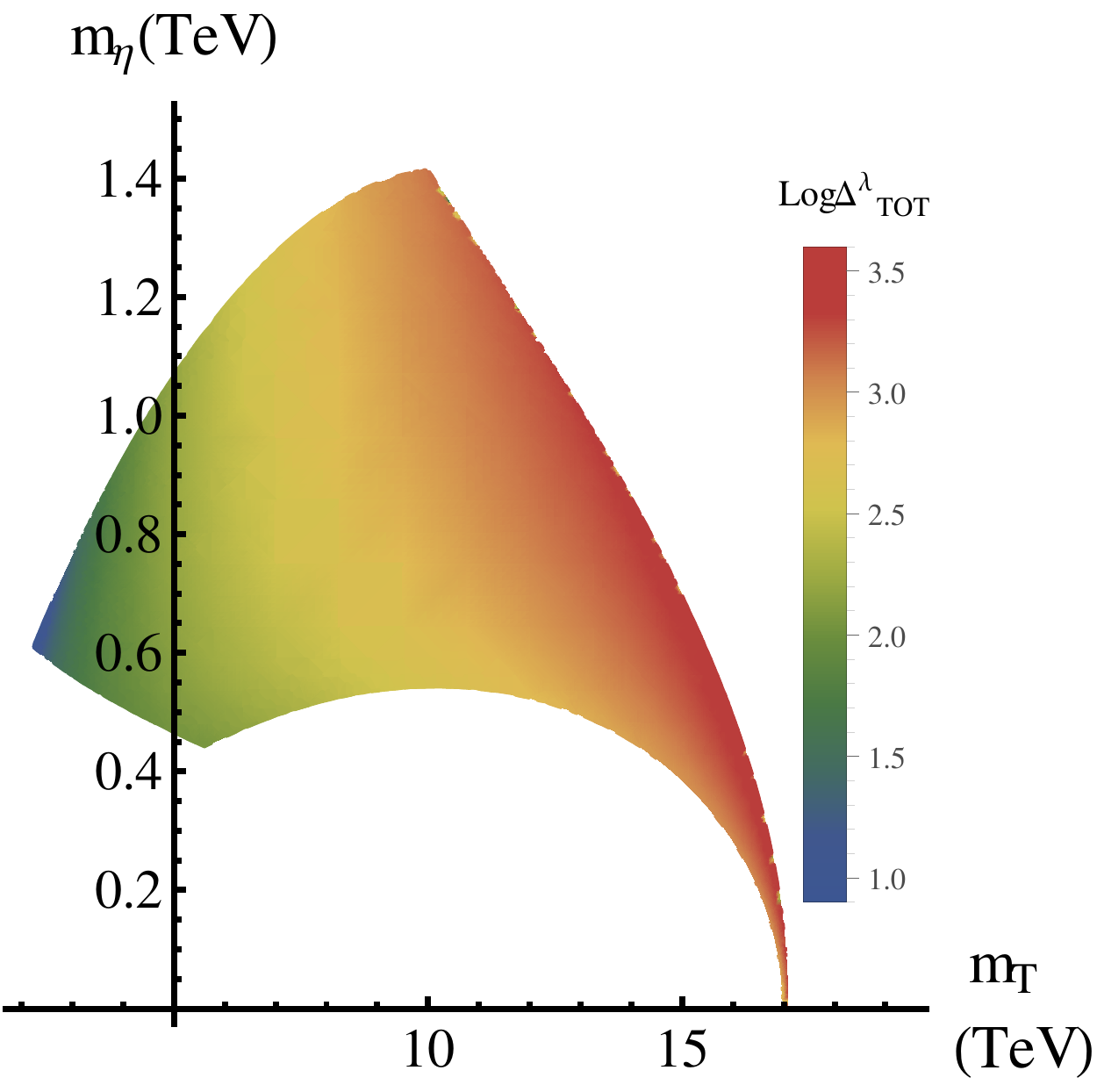}\quad\quad
\includegraphics[width=2.5in]{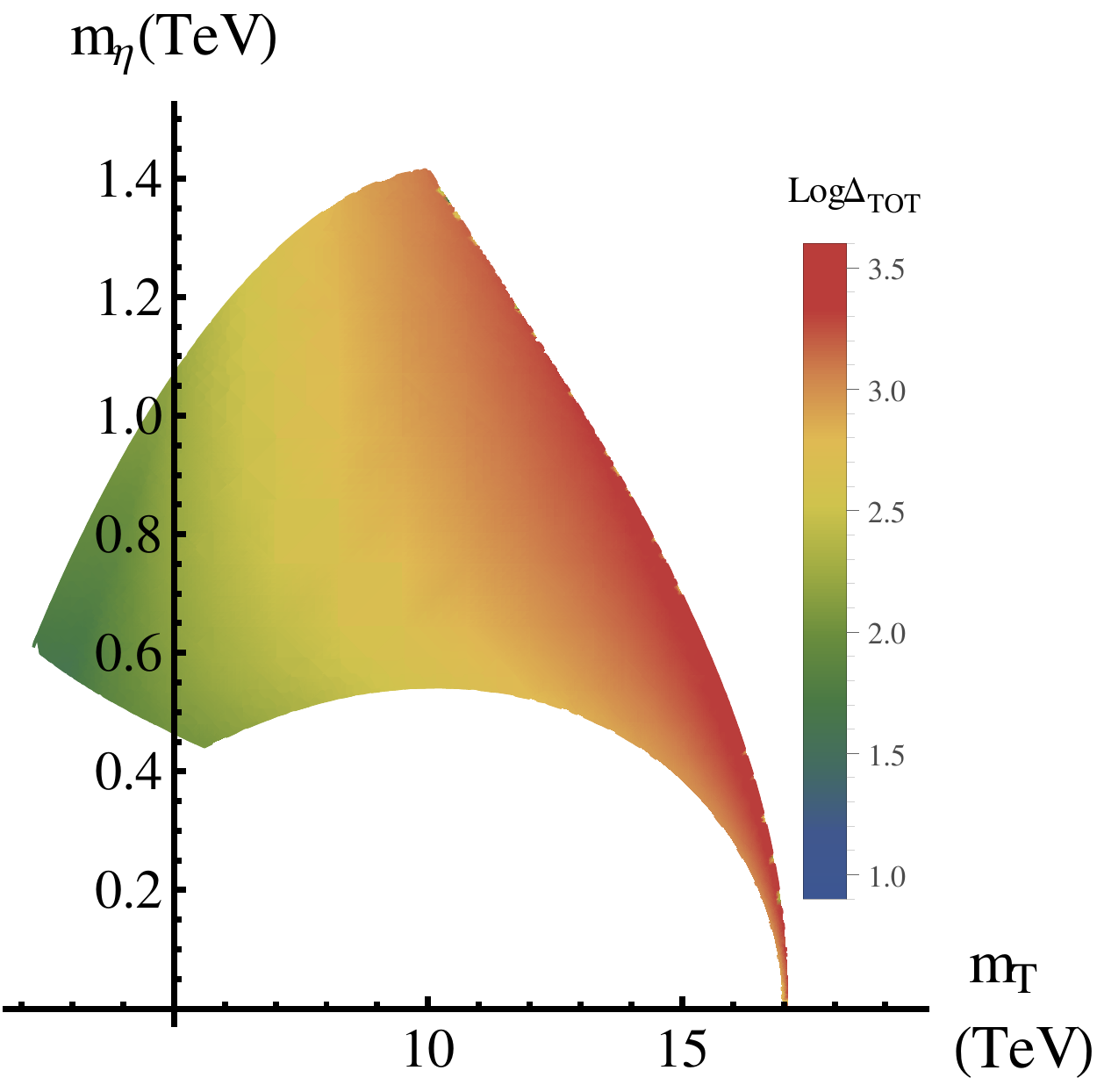}
\end{center}
\caption{\label{fig:dft10} Left: Density plot of $\text{Log}\Delta_\TOT^\lambda$ in the
$m_\eta-m_T$ plane for $f=10\TeV$. Right: Density plot of $\text{Log}\Delta_\TOT$ in the
$m_\eta-m_T$ plane for $f=10\TeV$. In both plots $\text{Log}$ means $\log_{10}$.}
\end{figure*}
In figure~\ref{fig:dft10} we present the density plot of $\text{Log}\Delta_\TOT^\lambda$
and $\text{Log}\Delta_\TOT$ in the $m_\eta-m_T$ plane for $f=10\TeV$. Two plots
are similar except for a small region in the leftmost corner. This is because
for $f=10\TeV$ $\Delta_\TOT$ is mainly determined by $\Delta_\TOT^\lambda$ except
for the region with small enough $m_T$ in which $\Delta_\TOT^\lambda<\Delta_\TOT^{\mu^2}$.
From the figure we see that for $f=10\TeV$ the parameter space favored by naturalness consideration
has small $m_T$ (down to $\sim 3\TeV$) and large $t_\beta$ (near the unitarity boundary),
with $\Delta_\TOT^{-1}$ approaching a few percent. The favored pseudo-axion mass is
around $600\GeV$ in this case. Unfortunately the parameter region corresponding to
a light pseduo-axion with mass smaller than $400\GeV$ is strongly disfavored by
naturalness, as can be inferred from the figure. Our results shown here to some extent
answers from the CEFT point of view the question in ref.~\cite{Dias:2007pv} which raises
concern about the fine-tuning in the SLH given a strong constraint on $f$.

\begin{figure*}[ht]
\begin{center}
\includegraphics[width=2.5in]{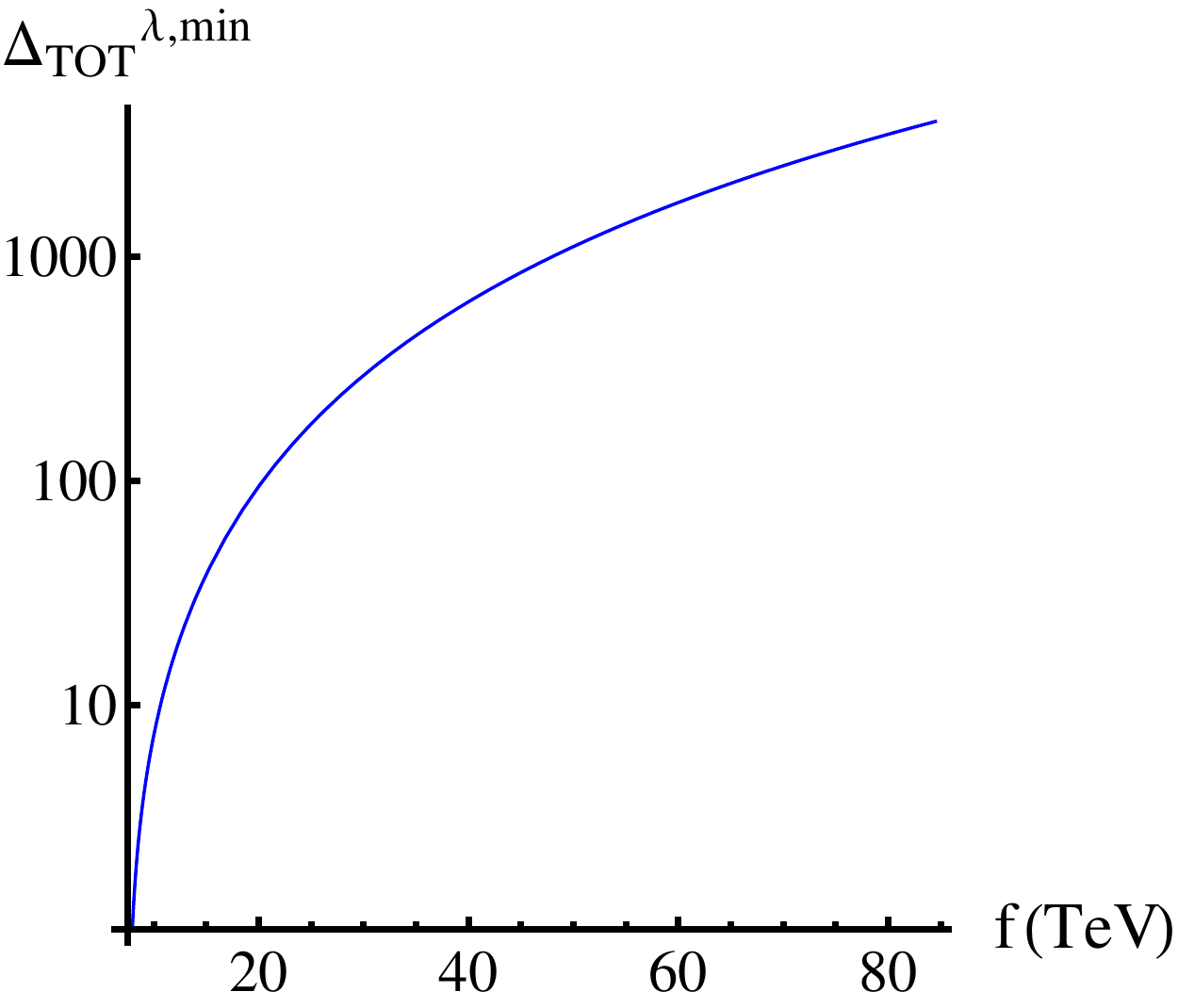}\quad\quad
\includegraphics[width=2.5in]{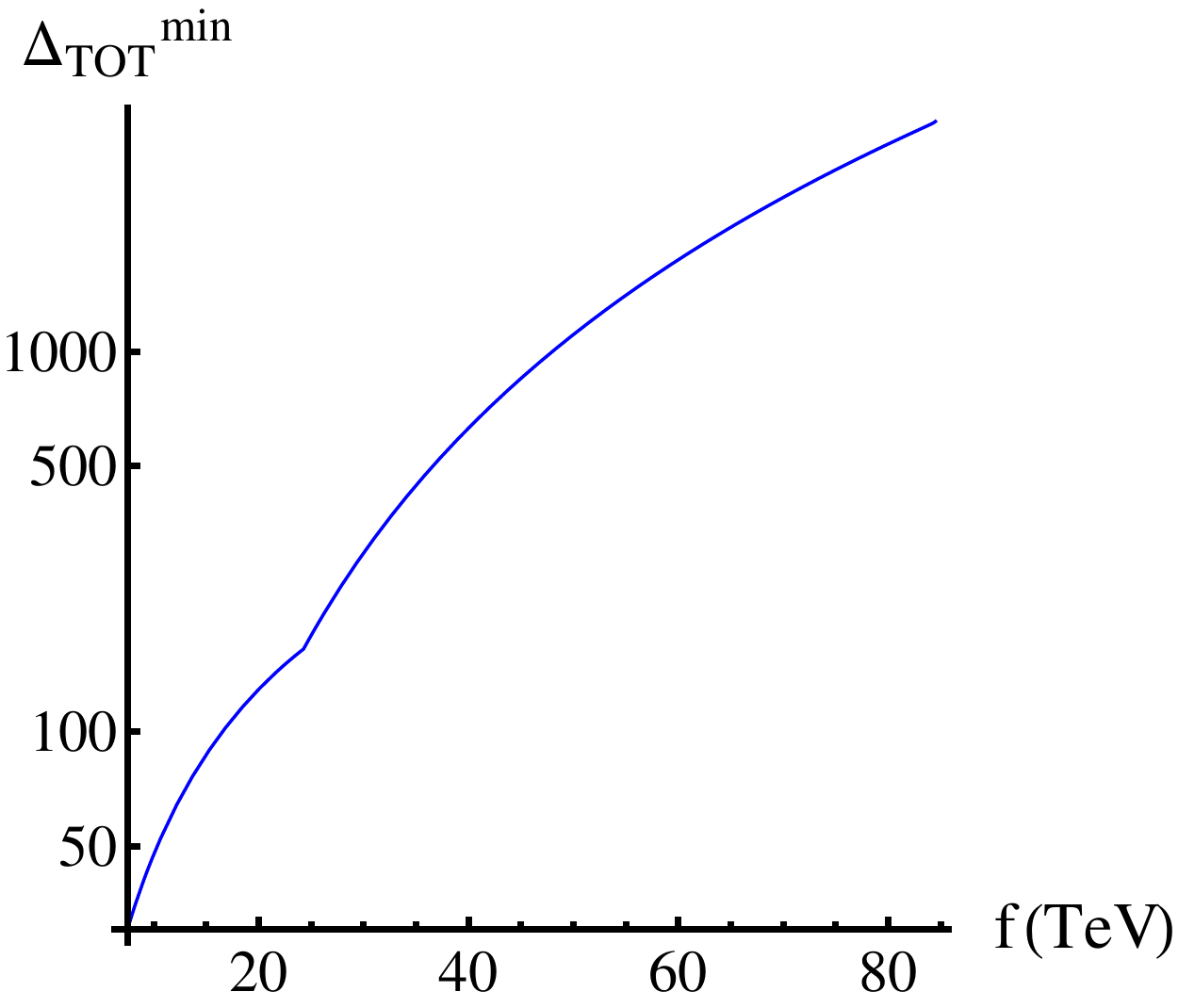}
\end{center}
\caption{\label{fig:mft1d} Left: Minimum $\Delta_\TOT^\lambda$ as a function of $f$.
Right: Minimum $\Delta_\TOT$ as a function of $f$.}
\end{figure*}
Figure~\ref{fig:mft1d} shows the minimum allowed $\Delta_\TOT^\lambda$
and $\Delta_\TOT$ as a function of $f$. It is not surprising that both of them
exhibit a monotonically increasing behavior. We note that for a sufficiently small
$f\sim 7.5\TeV$, $\Delta_\TOT^\lambda$ can reach zero, indicating that a vanishing
$\lambda_U$ at the unitarity cutoff is still allowed. However this value of $f$
is already near the boundary of LHC exclusion limit. For larger $f$,
$\Delta_\TOT^\lambda=0$ is then impossible which implies the NDCC assumption
with exactly vanishing $\lambda_U$ is no longer valid. The value of
minimum $\Delta_\TOT$ is determined by $\Delta_\TOT^{\mu^2}$ for $f<25\TeV$
and $\Delta_\TOT^\lambda$ for larger $f$. This yields a slight kink
at $f\sim 25\TeV$ in the right panel of figure~\ref{fig:mft1d}.

\section{Discussion and Conclusions}
\label{sec:dnc}

In this work we have analyzed the SLH scalar potential in an approach
consistent with the spirit of CEFT. The most important message we obtained
from the analysis is a mass relation connecting the pseudo-axion mass
and the top partner mass, Eq.~\eqref{eq:mr2}. The anti-correlation between
these two masses gives rise to interesting constraints on the SLH parameter
space. Especially, the minimally allowed $t_\beta$ increases with $f$
for $f>18\TeV$ (figure~\ref{fig:minmax}). On the other hand, the unitarity
constraint leads to a maximally allowed $t_\beta$ independent of $f$.
Therefore an absolute upper bound on $f$ ($f<85\TeV$) is obtained.
This in turn implies absolute upper bound on $Z'$ mass $m_{Z'}<48\TeV$
and top partner mass $m_T<19\TeV$. Pseudo-axion mass is also bounded
from above: $m_\eta$ can reach the maximum allowed value of
$m_\eta\approx 1.5\TeV$ when $f\approx 51\TeV$. We emphasized the importance
of retaining the field dependence in the field form factor $\Delta(\hat{h})$
for a quantitative analysis of the scalar potential, which is often ignored
by previous studies.

We have also analyzed the issue of Higgs mass naturalness in the SLH
in an approach consistent with the spirit of CEFT. The parameter space
with the least fine-tuning turns out to be characterized by a low value
of $f$ and a small $m_T$, with an inverse total tuning $\Delta_\TOT^{-1}$
at a few percent level. This can be achieved if $t_\beta$ is relatively
large, close to the unitarity upper bound, while predicting a pseudo-axion
mass at around $600\GeV$.

Although the pseudo-axion mass is bounded from above by a not-very-large
value of about $1.5\TeV$, it is nevertheless quite challenging to detect
such a particle at current and future colliders. The reason is that the
pseudo-axion couplings to SM particles are all suppressed by $\frac{v}{f}$,
and in some cases even by $\frac{v^3}{f^3}$ (such as the antisymmetric
$ZH\eta$ vertex~\cite{He:2017jjx}). Considering the current bound on $f$
($f>7.5\TeV$) the suppression is already quite significant. In such a situation
it is helpful to consider detecting $\eta$ from the decay of other new
heavy resonances such as $Z',T$~\cite{Kilian:2004pp,Cheung:2008zu}, other heavy quarks or leptons. The $Z'$
search using dilepton channel is expected to be most promising in terms of
discovery or most stringent in terms of constraint. Taking into account
off-shell $Z'$ contributions~\cite{Alioli:2017nzr} may even extend the reach and be helpful
to cover the very heavy region at a $100\TeV$ $pp$ collider. Another important target for
future collider search should be the top partner.
Although the heaviest mass value of about $19\TeV$ is beyond the reach of even
a $100\TeV$ $pp$ collider, the parameter region favored by naturalness consideration
should be well in reach.

A full program of testing the mass relation Eq.~\eqref{eq:mr2} is conceivable
if the associated mass scales are in reach of proposed future colliders. This should be
the case for a $100\TeV$ $pp$ collider if the SLH is realized in nature with $\Delta_\TOT^{-1}$ at a few
percent level. Four crucial quantities need to be measured, namely $f,t_\beta,m_\eta$
and $m_T$. $f$ can be determined once the $Z'$ is discovered and its mass measured
at, for example, the dilepton channel. For the region with small fine-tuning, the top
partner should be relatively light and therefore be within reach in the pair production
channel. The pseudo-axion $\eta$ might be discovered via $Z'$ or top partner decay.
Finally, some couplings related to the top partner and other heavy fermions are related
to $t_\beta$~\cite{Han:2005ru} and might provide a measurement thereof. Combining all
the information it would then be possible to understand or veto the EWSB as the consequence of
perturbative vacuum misalignment during the global symmetry breaking Eq.~\eqref{eq:gsb}.

It is certainly warranted to make a comparison between our CEFT-based approach to analyze the
scalar potential and the approach based on the NDCC assumption. An even sharper question
might be: what is the virtue of the CEFT-based approach compared to using a ``floating
cutoff'' in previous analyses? We may criticize an analysis using the NDCC assumption
with a fixed cutoff value by noting that any of its EWSB predictions which depend
on the choice of the cutoff have an uncertainty which is hard to quantify. However, if
a ``floating cutoff'' is used instead of a fixed value, then it seems that it would
yield the same parameter space as compared to our CEFT-based approach. Our answer to this
question consists of two aspects. First, it is always desirable that physics theory
be not only a tool for calculation, but also can be interpreted in a conceptually
consistent manner. In this regard, the practice of imparting physical meaning to
the regularization cutoff is not conceptually solid. For example, if we consider
an asymptotically free theory like QCD, then we would realize that its perturbative calculation
involves logarithmic divergences which need not be cut off by any new particles
since QCD can be consistently extrapolated to arbitrarily high energy, yet the
divergences are actually harmless~\cite{Peskin:1995ev}. Another example is that
imparting physical meaning to the regularization cutoff could lead to violation
of field redefinition invariance, as pointed out in ref.~\cite{Burgess:1992gx}.
On the other hand, the CEFT-based approach is based on standard renormalization
theory and is therefore conceptually clear. Second, from a pratical point of view,
the CEFT approach makes manifest the reliable prediction of the theory. In the SLH
case, it is the mass relation which has been hidden for a long time under
previous studies. It is also worth noting that the CEFT-based approach facilitates
the analysis of physical fine-tuning, as demonstrated in Section~\ref{sec:nislh}.

The CEFT-based approach to scalar potential and fine-tuning analysis adopted in
this work can be carried over to a wide class of Little Higgs and Twin Higgs models
as well. The expected outcome should contain at least various mass relations
which pertain to each of the model under investigation. These mass relations should
serve as crucial tests of the related EWSB mechanism. The parameter space with
small fine-tuning is expected to be associated with a small top partner mass.
In conjunction with other theoretical considerations (e.g. perturbative unitarity and
vacuum stability) and experimental probes, it is hopeful that we may obtain
a deeper understanding of the EWSB, one of the most important pillars of
contemporary particle physics.

\acknowledgments

We thank Hsin-Chia Cheng and Wai-Yee Keung for helpful discussion.
This work was supported in part by the Natural Science
Foundation of China (Grants No. 11135003, No. 11375014
and No. 11635001), the China Postdoctoral Science
Foundation (Grant No. 2017M610992) and the MoST of Taiwan under
the grant no.:105-2112-M-007-028-MY3.


\bibliography{slhspa_v4}
\bibliographystyle{JHEP}

\end{document}